\documentclass[fleqn,usenatbib]{mnras}

\usepackage{newtxtext,newtxmath}
\usepackage[T1]{fontenc}
\usepackage{amsmath}    
\usepackage{mathtools}
\usepackage{comment}
\usepackage{subfigure}
\usepackage{graphicx}
\usepackage{lipsum}
\usepackage{braket}
\usepackage{color}
\usepackage{algorithm}
\usepackage[noend]{algpseudocode}
\usepackage[dvipsnames]{xcolor}
\usepackage{float}
\usepackage{gensymb}
\usepackage{siunitx}

\newcommand{\flag}[1]{\texttt{\lowercase{#1}}}
\newcommand{\FLAG}[1]{\texttt{\uppercase{#1}}}
\newcommand{\Flag}[1]{\texttt{#1}}

\newcommand{\relensing}{\flag{relensing}}
\newcommand{\python}{\flag{python}}
\newcommand{\glafic}{\flag{glafic}}
\newcommand{\numpy}{\flag{numpy}}
\newcommand{\scipy}{\flag{scipy}}
\newcommand{\matplotlib}{\flag{matplotlib}}
\newcommand{\numba}{\flag{numba}}
\newcommand{\astropy}{\flag{astropy}}
\newcommand{\github}{\texttt{GitHub}}

\newcommand{\LensPerfect}{\Flag{LensPerfect}}
\newcommand{\SWUnited}{\Flag{SWUnited}}
\newcommand{\SaWLens}{\Flag{SaWLens(2)}}

\newcommand{\LENSTOOL}{\FLAG{LENSTOOL}}
\newcommand{\LTM}{\FLAG{LTM}}
\newcommand{\GRALE}{\FLAG{GRALE}}
\newcommand{\WSLAP}{\FLAG{WSLAP}}
\newcommand{\WSLAPP}{\FLAG{WSLAP+}}

\DeclareRobustCommand{\VAN}[3]{#2}
\let\VANthebibliography\thebibliography
\def\thebibliography{\DeclareRobustCommand{\VAN}[3]{##3}\VANthebibliography}

\title{\relensing: Reconstructing the mass profile of galaxy clusters from gravitational lensing}

\author[D. A. Torres-Ballesteros and L. Casta\~neda]{
Daniel A. Torres-Ballesteros,$^{1}$\thanks{E-mail:daatorresba@unal.edu.co} and Leonardo Casta\~neda$^{1}$\thanks{E-mail: lcastanedac@unal.edu.co}
\\
$^{1}$Observatorio Astron\'omico Nacional, Universidad Nacional de Colombia, Carrera 30 Calle 45-03, P.A. 111321 Bogot\'a, Colombia
}

\date{Accepted XXX. Received YYY; in original form ZZZ}

\pubyear{2021}

\begin{document}
\label{firstpage}
\pagerange{\pageref{firstpage}--\pageref{lastpage}}
\maketitle

\begin{abstract}
In this work we present \relensing, a package written in \python\ whose goal is to model galaxy clusters from gravitational lensing. With \relensing\ we extend the amount of software available, which provides the scientific community with a wide range of models that help to compare and therefore validate the physical results that rely on them. We implement a free-form approach which computes the gravitational deflection potential on an adaptive irregular grid, from which one can characterize the cluster and its properties as a gravitational lens. Here, we use two alternative penalty functions to constrain strong lensing. We apply \relensing\ to two toy models, in order to explore under which conditions one can get a better performance in the reconstruction. We find that by applying a smoothing to the deflection potential, we are able to increase the capability of this approach to recover the shape and size of the mass profile of galaxy clusters, as well as its magnification map. This translates into a better estimation of the critical and caustic curves. The power that the smoothing provides is also tested on the simulated clusters Ares and Hera, for which we get an rms on the lens plane of $\sim 0.17\,arcsec$ and $\sim 0.16\,arcsec$, respectively. Our results represent an improvement with respect to reconstructions that were carried out with methods of the same nature as \relensing. At the same time, the smoothing also increases the stability of our implementation, and decreases the computation time. In its current state, \relensing\ is available upon request.
\end{abstract}




\section{Introduction}\label{sec: introductiuon}

Gravitational lensing has proven to be a powerful tool for understanding the universe at different scales (e.g. \cite{blandford1992cosmo, schneider2006gravitational}), where its power relies upon its simplicity, since it only depends on the mass distribution that acts as lens, irrespective of its nature (baryonic or not) or its dynamical state. 

It is of particular interest its application to the study of galaxy clusters and their mass distribution. As the most massive objects in the universe, galaxy clusters provide a framework for cosmological probes, such as testing the underlying gravitational theory (e.g. \cite{lam2012gravitytest,pizzuti2016gravitytest, cataneo2018gravitytest}) and cosmological model (e.g. \cite{ gilmore2009cosmology,daloisio2011cosmology, jullo2010cosmological}), and also provide strong constrains to the paradigm of structure formation and evolution (e.g see \cite{allen2011cosmological, kravtsov2012clusterformation} for reviews). In addition, galaxy clusters also act as effective cosmic telescopes, since their angular size and mass turn them into the biggest and strongest manifestation of gravitational lensing in the universe. Therefore, galaxy clusters are capable of magnifying faint background sources that would otherwise remain unseen, or for which it would be difficult to infer any significant information. This is for example the case for galaxies with high redshift (the first galaxies ever formed in the universe) (e.g. \cite{kneib2004cosmictelescope,bradac2009cosmictelescope, coe2013cosmictelescope, richard2014cosmictelescope, oesch2015cosmictelescope, mcleond2016cosmictelescope, furtak2021cosmictelescope}). For a deeper description of the role that gravitation lensing plays in the context of galaxy clusters refer to e.g. \cite{kneib2011gcreview, hoekstra2013gcreview, umetsu2020review} and the references therein.

The problem of making an estimation of the mass profile of a galaxy cluster from gravitational lensing has been approached from different perspectives over the years. As a result, several methods have been developed. In general terms, they are usually classified as parametric or free-form (sometimes also labeled as non-parametric) methods. Parametric methods are characterized by employing an input model that depends on a set of parameters which needs to reproduce the input data as well as possible. This can be done for example, by means of Bayesian methods such as MCMC or Nested Sampling. The input model can be for example, the mass distribution composed of a superposition of baryonic and dark matter components for the different cluster members, and which are chosen taking into account certain physical and geometrical properties that are expected to satisfy the properties shown by the galaxy cluster. On the other hand, free-form methods generally work on a grid on which some quantity of interest is defined, for example the convergence or the deflection potential, and they need to be found. Therefore, free-form methods are parametric on the inside, since there is a set of parameters that need to be found in order to achieve the reconstruction, but they are different to pure parametric methods in the sense that does not exist an imposed input model that constrains the reconstruction. 

Among parametric approaches we can find e.g. \LENSTOOL\ \cite{kneib1996lenstool, jullo2007bayesian, newman2013lenstool, schafer2020lenstool}, Light-Traces-Mass (\LTM) \cite{broadhurst2005ltm}, and \glafic\ \cite{oguri2010glafic}, Meanwhile, for free-form approaches we have e.g. \cite{bartelmann1996freeform}, \cite{abdelsalam1998a370}, \cite{saha1997freeform}, \cite{seitz1998entropy}, \cite{cacciato2006freeform}, \cite{jee2007freeform}, \cite{deb2008reconstruction}, \GRALE\ \cite{liesenborgs2006grale, liesenborgs2007grale,liesenborgs2009grale, liesenborgs2020gralewl, ghosh2020grale}, \WSLAP\ \cite{diego2005wslap, diego2007wslap}, \SWUnited\ \cite{bradac2005reco, bradac2005reco2, bradac2009cosmictelescope}, \LensPerfect\ \cite{coe2008lp}, and \SaWLens\ \cite{merten2009reco,merten2011reco2, merten2016freemesh, huber2019sw+x}. There are also available hybrid methods that combine elements of both approaches e.g. \cite{jullo2009lenstoolhybrid, niemiec2020lenstoolhybrid} (which are part of \LENSTOOL), and \WSLAPP\ \cite{sendra2014wslap+, diego2016wslap+} (an extension of \WSLAP).

Nowadays, the relevant observations in this regard are open access, as it is the case for Hubble Frontier Fields (HFF) \cite{lotz2017hff}, Reionization Lensing Cluster Survey (RELICS) \cite{coe2019relics}, Cluster Lensing and Supernova 
Survey with Hubble (CLASH) \cite{postman2012clash}. Additionally, there is an expected increase in both quality and quantity of observations from upcoming ground/space based observatories such as James Webb Space Telescope (JWST) \cite{gardner2006jwst}, Euclid \cite{laureijs2009euclid, laureijs2011euclid}, and Vera C. Rubin Observatory (LSST) \cite{ivezic2019lsst}. Therefore, it is pertinent to have access to a variety of open access software that aims to model galaxy clusters from gravitational lensing, in order to be able to compare, and validate those models, as well as the physical results that rely on them (e.g \cite{priewe2017HFFcomparison, meneghetti2017HFFcomparison, strait2018abell370, remolina2018macsj0416.1-2403}).

Therefore, in this work we describe what is behind \relensing, an easy-to-use package currently written in \python, in which a free-form reconstruction method is implemented. The method that we discuss here is based upon the work previously presented by \cite{bradac2005reco, bradac2009cosmictelescope}, which itself is an extension of \cite{bartelmann1996freeform}.

We explore the effectiveness of an alternative finite difference approach. Also, we extend the grid refinement to an adaptive irregular one, that is intended to follow the observational data from strong lensing as well as the shape of the main deflectors.  Additionally, we discuss two different approaches to the strong lensing penalty function, which along with the introduction of a smoothing process enhance the performance of this approach, reducing the computation time and providing a more accurate reconstruction.

This paper is organized as follows. In Sec. \ref{sec: lensig} we discuss the relevant aspects of gravitational lensing theory, and also introduce the notation used throughout the paper. Then, in Sec. \ref{sec: reconstruction method} we describe the reconstruction method in detail. Once the reconstruction framework is settled, in Sec. \ref{sec: simple models} we apply the reconstruction to two simple distributions, in order to explore how \relensing\ behaves and how to get the best out of it. In Sec. \ref{sec: realistic distributions} we apply the reconstruction to Ares and Hera (see \cite{meneghetti2017HFFcomparison}), which provide a way of testing the performance of \relensing\ in a more realistic scenario. Finally, in Sec. \ref{sec: conclusions} we present the conclusions of this work.

For the mock catalogues discussed in Sec. \ref{sec: mock catalogues}, and the reconstructions presented in this work, we consider a flat $\Lambda$CDM cosmology with density parameter $\Omega_{m,0} = 0.3$ and Hubble constant $H_0=70\,km\,s^{-1}\,Mpc^{-1}$.

\section{Gravitational lensing notation}\label{sec: lensig}
In this work we focus on the gravitational lensing effect under the thin lens approximation. A presentation of this topic deeper than the one intended here, can be found, for example in \cite{schneider1992lenses, schneider2006gravitational}. 

In this context, the galaxy cluster with redshift $z_l$ takes the place of the lens, and it is characterized by the projected mass density $\Sigma = \Sigma(\boldsymbol{\theta})$, with $\boldsymbol{\theta}$ being an angular position from the observer perspective. Then, the aim is to recover $\Sigma$ somehow. For this task, we need to compare observations to theory. Therefore, as observables we will make use of multiply imaged systems (strong regime) (e.g. \cite{hattori1999strong,schneider2006gravitational}), since galaxy clusters  are rich in those systems, and we will also consider the ellipticity of weakly distorted and singly imaged  background sources (weak regime) (e.g. \cite{bartelmann2001weak, schneider2006gravitational, Hoekstra2013weaklecture,mandelbaum2018weakreview,umetsu2020review}).

The aforementioned observables can be characterized by means of the convergence $\kappa =\kappa(z, \boldsymbol{\theta})$, the shear $\gamma = \gamma(z,\boldsymbol{\theta})$, and also the reduced (or scaled) deflection angle $\boldsymbol{\alpha} = \boldsymbol{\alpha}(z,\boldsymbol{\theta})$ (hereafter we will refer to it as deflection angle for simplicity), which are defined in terms of potential $\psi = \psi(z,\boldsymbol{\theta})$. Here $\kappa$ is a direct estimator of the mass profile, since it is defined as
\begin{equation}\label{eq: convergence}
\kappa := \frac{\Sigma}{\Sigma_{cr}} = \dfrac{1}{2}\nabla^2 \psi \quad\text{with}\quad\Sigma_{cr}:=\frac{c^2 D(z)}{4\pi G D(z_l)D(z_l, z)},
\end{equation}
where $\Sigma_{cr}=\Sigma_{cr}(z,z_l)$ is known as critic mass density, and it is defined in terms of the angular diameter distances from the observer to the lens $D(z_l)$, from the observer to source $D(z)$ of interest, and finally from the lens to such source $D(z_l,z)$. Thus, we refer to $\kappa$ as the mass profile. Similarly, $\gamma$ is a dimensionless complex quantity whose components are
\begin{equation}\label{eq: shear components}
\gamma_1 :=\frac{1}{2}\Bigg(\frac{\partial^2\psi}{\partial\theta_1^2} -  \frac{\partial^2\psi}{\partial\theta_2^2}\Bigg)\quad\text{and}\quad \gamma_2:=\frac{\partial^2\psi}{\partial\theta_1\partial\theta_2},
\end{equation}
with $\gamma:= \gamma_1 + i\gamma_2$. In addition, $\boldsymbol{\alpha}$ reads
\begin{equation}\label{eq: deflectin angle}
\boldsymbol{\alpha}:=\nabla \psi.    
\end{equation}

From \eqref{eq: convergence} it is clear that once $\psi$ is known, the mass profile is recovered. Now, since $\psi$ depends on $z$ it is convenient to define a reference redshift $z_r>z_l$, so that given $\psi_r$ one can write $\psi$, $\kappa$. $\gamma$ and $\boldsymbol{\alpha}$ in terms of $\psi_r$ as well as in terms of $\kappa_r$, $\gamma_r$ and $\boldsymbol{\alpha}_r$, respectively. This transition is done by using the cosmological weight $Z=Z(z,z_r,z_l)$ defined as
\begin{equation}\label{eq: cosmological weight}
Z:=\frac{D(z_l,z)D(z_r)}{D(z)D(z_l,z_r)}U(z),
\end{equation}
with
\begin{equation}\label{eq: step}
U(z)=\begin{cases}
\begin{array}{c}
1\\
0
\end{array} & \begin{array}{l}
\text{if}\quad z>z_l\\
\text{if}\quad z\leq z_l
\end{array}\end{cases},  
\end{equation}
from which one gets $\psi = Z\psi_r$, $\kappa=Z\kappa_r$, $\gamma = Z\gamma_r$ and $\boldsymbol{\alpha} = Z\boldsymbol{\alpha}_r$. In  this way, we only need to deal with $\psi_r$, which simplifies the problem significantly. 

With respect to the observables, in the strong regime the multiply imaged systems can in principle be characterized by solving the lens equation 
\begin{equation}\label{eq: lens equation}
\boldsymbol{\beta} = \boldsymbol{\theta} - Z\boldsymbol{\alpha}_r,
\end{equation}
which relates the true angular position $\boldsymbol{\beta}$ to the observed angular position $\boldsymbol{\theta}$. Alternatively, it is simpler to work on the source plane, taking into account that the multiple images of a given system have to converge to a unique source.

On the other hand, with respect to the weak regime, distortions (shape, size and orientation) at this level are characterized by means of source ellipticity. Among the different definitions of ellipticity (e.g. \cite{schneider2006gravitational}), the more suited for the task in hand is the complex ellipticity $\epsilon = \epsilon(z, \boldsymbol{\theta})$, which is related to the source intrinsic or true ellipticity $\epsilon_s$ by means of the expression 
\begin{equation}\label{eq: ellipticiy}
\epsilon=\begin{cases}
\begin{array}{c}
\dfrac{\epsilon_s+g}{1+g^{*}\epsilon_s}\\ \\
\dfrac{1+g\epsilon_s^*}{\epsilon^{*}_s+g^{*}}
\end{array} & \begin{array}{l}
\text{if}\quad |g|\leq 1\\ \\
\text{if}\quad |g|>1
\end{array}\end{cases},
\end{equation}
where the asterisk stands for complex conjugate, and the reduced shear $g=g(z)$ is defined as
\begin{equation}\label{eq: reduced shear}
g=\frac{Z\gamma_r}{1-Z\kappa_r}.
\end{equation}

Here $\epsilon_s$ is not an observable, therefore \eqref{eq: ellipticiy} cannot be used directly. This problem is avoided by considering the average intrinsic ellipticity for several (as many as possible) sources in a local neighborhood to be $\braket{\epsilon_s}=0$. As a result, for a given redshift the average observed ellipticity reads
\begin{equation}\label{eq: ellipticiy average}
\braket{\epsilon}=\begin{cases}
\begin{array}{c}
g\\ \\
\dfrac{1}{g^*}
\end{array} & \begin{array}{l}
\text{if}\quad |g|\leq 1\\ \\
\text{if}\quad |g|>1
\end{array}\end{cases},
\end{equation}
yielding a direct estimation of the local reduced shear, which allows us to directly contrast observations to theory at weak regime. Further details can be found in \cite{schneider1995weak1, seitz1997weak3}.

\section{Reconstruction method}\label{sec: reconstruction method}
The aim of those methods derived from \cite{bartelmann1996freeform} and its extensions is to constrain the deflection potential $\psi$ in a given grid by making use of gravitational lensing observations. No assumption about mass traces light is made.

Consider (in general) an irregular grid defined on the lens plane, which is composed of $N$ nodes whose angular position is $\boldsymbol{\theta}_j$ (with $j=1,2,\dots,N$). On each node a deflection potential $\psi_j$ is assigned, and such potentials are constrained from the observables described in Sec. \ref{sec: lensig}. To do so, we implement the penalty function
\begin{equation}
\label{eq: penalty function}
\chi^2(\psi_j):=\chi^2_s(\psi_j)+\chi^2_w(\psi_j)+ \chi^2_{\kappa(R)}(\psi_j) + \chi^2_{\gamma(R)}(\psi_j),
\end{equation}
where $\chi^2_s$ constrains the strong regime, $\chi^2_w$ constrains the weak regime, and both $\chi^2_{\kappa(R)}$ and $\chi^2_{\gamma(R)}$ are regularization terms. Here, \eqref{eq: penalty function} needs to be explicitly written in terms of the $N$ deflection potentials, since they are the parameters to be found. This task can be achieved by minimizing \eqref{eq: penalty function} with respect to every $\psi_k$ (with $k=1,2,\dots,N$), which leads to $N$ equations of the form
\begin{equation}\label{eq: minimization}
\frac{\partial\chi^2_w}{\partial\psi_k}+\frac{\partial\chi^2_s}{\partial\psi_k}+\frac{\partial \chi^2_{\kappa(R)}}{\partial \psi_k}+\frac{\partial \chi^2_{\gamma(R)}}{\partial \psi_{k}}=0,
\end{equation}
each of them with $N$ unknowns. The system of equations is in general nonlinear, making it difficult to solve. Linearizing (if possible) such system simplifies the implementation and reduces the computation time, which is an advantage in practical terms, considering that reliable results are obtained. In Sec. \ref{sec: penalty function} we explain how to perform such linearization.

Now, from Sec. \ref{sec: lensig} it is clear that the observables we are interested in can be characterized by $\kappa$, $\gamma$, and $\boldsymbol{\alpha}$, which are defined in terms of either first or second order derivatives of $\psi$. Since we are working on a grid, the most direct way to explicitly write these quantities in terms of $\psi$ is by using finite differences. In this context, the advantage of finite differences over other methods lies in its  capability to express the derivatives (of any order) of $\psi$ as a linear combination of $\psi_j$ (defined on the grid), and it also makes the linearization of \eqref{eq: minimization} simple. In this direction, $\kappa$, $\gamma$, $\boldsymbol{\alpha}$, and $\psi$ itself, evaluated at an arbitrary angular position $\boldsymbol{\theta_i}$ (it does not need to correspond to a node) can be written in terms of every $\psi_j$ as
\begin{equation}
\label{eq: finite diff pot}
\psi_i=\sum_{j=1}^{N}\mathcal{P}_{ij}\psi_j,
\end{equation}
\begin{equation}
\label{eq: finite diff con}
\kappa_i=\sum_{j=1}^{N}\mathcal{K}_{ij}\psi_j,
\end{equation}
\begin{equation}
\label{eq: finite diff gam}
\gamma_{n,i}=\sum_{j=1}^{N}\mathcal{G}_{n,ij}\psi_j,
\end{equation}
\begin{equation}
\label{eq: finite diff alp}
\alpha_{n,i}=\sum_{j=1}^{N}\mathcal{D}_{n,ij}\psi_j,
\end{equation}
with $n=1,2$ representing the components of $\gamma$ and $\boldsymbol{\alpha}$, respectively. The coefficients (or nodes weight) $\mathcal{P}_{ij}$, $\mathcal{K}_{ij}$, $\mathcal{G}_{n,ij}$ and $\mathcal{D}_{n,ij}$ depend on what form the grid has, i.e. if it is regular (e.g. \cite{bradac2005reco,deb2008reconstruction, merten2009reco}) or irregular (e.g. \cite{bradac2009cosmictelescope, merten2016freemesh}), and also on how finite differences and the interpolation are carried out. In Sec.
\ref{sec: finite difference} we describe how we obtain such weights. Since $\boldsymbol{\theta}_i$ is arbitrary (within the region of interest), equations \eqref{eq: finite diff pot} - \eqref{eq: finite diff alp} also work to interpolate.

Once the different penalty functions in \eqref{eq: penalty function} are explicitly  written in terms of $\psi_j$, we have that their contribution to \eqref{eq: minimization} after the corresponding linearization yields an expression of the form
\begin{equation}
\label{eq: minimization components}
\frac{\partial\chi^2_p}{\partial\psi_k}=2\sum_{j=1}^{N}\mathcal{W}_{kj}^{(p)}\psi_{j}-2\mathcal{V}_{k}^{(p)},
\end{equation}
where $p=s$ stands for strong regime, $p=w$ for weak regime, and $p=\kappa$ and $p=\gamma$ stand for both regularization terms. In Sec. \ref{sec: penalty function} we describe how $\mathcal{W}_{kj}^{(p)}$ and $\mathcal{V}_{k}^{(p)}$ are defined in each case. From \eqref{eq: minimization components} it is clear that $\eqref{eq: minimization}$ reduces to 
\begin{equation}
\label{eq: system of linear equations}
\sum_{j=1}^{N}\mathcal{W}_{kj}\psi_j=\mathcal{V}_{k},
\end{equation} 
where $\mathcal{W}_{kj} = \mathcal{W}_{kj}^{(w)}+\mathcal{W}_{kj}^{(s)}+\mathcal{W}_{kj}^{(\kappa)}+\mathcal{W}_{kj}^{(\gamma)}$ and $\mathcal{V}_{k} = \mathcal{V}_{k}^{(w)}+\mathcal{V}_{k}^{(s)}+\mathcal{V}_{k}^{(\kappa)}+\mathcal{V}_{k}^{(\gamma)}$. Here the coefficient $\mathcal{W}_{kj}$ represent the $j$-th node weight for the $k$-th linear equation. 

The deflection potential admits two transformations that introduce degrees of freedom in the reconstruction, which may affect different aspects of the results. On one hand, transformation
\begin{equation}\label{eq: potential transformation sourceplane}
\psi\to\psi^{\prime}=\psi+\psi_0+\boldsymbol{c}\cdot\boldsymbol{\theta}
\end{equation}
introduces three degrees of freedom, one through the constant term $\psi_0$, and two from the bidimensional constant vector $\boldsymbol{c}$. On the other hand, transformation
\begin{equation}\label{eq: potential transformation masssheet}
\psi \to \psi^{\prime}= (1-\lambda)|\boldsymbol{\theta}|^2/2 + \lambda\psi,
\end{equation}
responsible for the well known Mass Sheet Degeneracy (MSD for short), introduces an extra degree of freedom through the constant term $\lambda$. Following \cite{bradac2005reco}, we have that such degrees of freedom can be fixed by keeping the deflection potential fixed on one node for each degree of freedom. Since we have four degrees of freedom, we will hold $\psi_j=0$ fixed on each of the four corners of the grid if both transformations are relevant, or only on three of the four corners of the grid if the MSD in not considered to be a concern. This leaves us respectively with a $(N-4)\times(N-4)$ system of linear equations to be solved.

Observations of galaxy clusters have shown the existence of several systems of multiple images, and for many of them the redshift is known; at least up to some extent. When such redshifts are sufficiently diverse, in principle the MSD breaks. Under these conditions the MSD is not a concern, so that we are left only with three degrees of freedom from \eqref{eq: potential transformation sourceplane}. Thus, in this case we are left with a $(N-3)\times(N-3)$ system of linear equations. In this work this condition is met. Hence, as well as in \cite{bradac2005reco}, we will not consider the MSD to be a problem, which is supported by the results that we get.

Now, before we describe which finite difference approach that we use, let us continue describing, in general, how the reconstruction method works. At this point, one might think that solving \eqref{eq: system of linear equations} is just enough for getting the reconstruction done. Unfortunately, this leads to a low resolution and under reconstructed profile. In this regard, \cite{bradac2005reco} found that two levels of iterations solve the problem. Therefore, keeping their notation, we will refer to them as the inner and outer levels. The inner level is in charge of computing the deflection potential iteratively, while the outer level is responsible for producing the grid refinement that produce a finer and adapted grid on which a new estimation of $\psi$ takes place.

In general terms, a description of how this method works is presented in Algorithm \ref{algo: reco}.
\begin{algorithm}[!hbt]
\caption{Reconstruction}
\label{algo: reco}
\begin{algorithmic}[1]
\State{\textbf{Input}: Initial conditions and the initial guess $\psi^{(0)}$.\label{algo: input}}
\State{\label{algo: cero}{Compute $\kappa^{(0)}$, $\gamma_1^{(0)}$ and $\gamma_2^{(0)}$ from $\psi^{(0)}$.}}
\While{\label{algo: outer} outer\_end = False} 
\State{\label{algo: coeff} Compute the nodes weight.}
\While{\label{algo: inner} inner\_end = False}
\State{\label{algo: sle} Compute the system of linear equations needed and solve it $\longrightarrow \psi^{(n)}$.}
\State{\label{algo: pot_sm} Smoothing of $\psi^{(n)}$ $\longrightarrow$ New 
$\psi^{(n)}$ (optional).}
\State{\label{algo: kappa} Compute $\kappa^{(n)}$ from $\psi^{(n)}$.\vspace{1.5mm}}
\If{\label{algo: converge}$\big\lvert\kappa^{(n)}_{j}-\kappa^{(n-1)}_{j}\big\rvert\leq \text{tolerance}$\vspace{1.5mm}} 
\State{\label{algo: inner end}inner\_end = True.} 
\EndIf
\EndWhile
\State{\label{algo: penalty} Compute $\chi^2_{s}$.}
\If{\label{algo: break} $\chi^2_s \leq q(2N_{\text{img}})$\vspace{1mm}}
\State{\label{algo: outer end}outer\_end = True.}
\EndIf
\If{\label{algo: outer new} outer\_end = False}
\State{\label{algo: refinement} Apply grid refinement.}
\State{\label{algo: new value} Compute $\psi^{(m)}$, $\kappa^{(m)}$, $\gamma_1^{(m)}$ and $\gamma_2^{(m)}$ on the new grid.}
\EndIf
\EndWhile
\State{\label{algo: return}\Return{$\psi$}}
\end{algorithmic}
\end{algorithm}

This algorithm has as input  (line \ref{algo: input}) a set of parameters that are used along the reconstruction. Among them, we have the grid initial and refinement conditions (more details in Sec. \ref{sec: grid refinement}), regularization weights $\eta_{\kappa}$ and $\eta_{\gamma}$ (more details in Sec. \ref{sec: penalty function regularization}), whether constrains from both strong and weak regimes are used or only from the strong regime, and the type of $\chi^2_s$ that is going to be used (more details in Sec. \ref{sec: penalty function strong}). Last but not least, as input we have the initial guess $\psi^{(0)}$ for the deflection potential, from which we compute $\kappa^{(0)}$, $\gamma_1^{(0)}$ and $\gamma_2^{(0)}$ (line \ref{algo: cero} ). These quantities are needed to initialize the reconstruction (more details in Sec. \ref{sec: penalty function}).

Once we have set the initial conditions, the outer level starts (line \ref{algo: outer}). At the beginning of each outer iteration we compute the node weights for every node and data point (line \ref{algo: coeff}). Such weights come from the finite difference scheme and the interpolation (when required), and have to be computed once for each outer iteration. With these  weights in hand, the inner level starts (line \ref{algo: inner}).

The inner level is in charge of computing the system of linear equations \eqref{eq: system of linear equations} and its solution, so that we get $\psi^{(n)}$; a new version of the deflection potential. Here, $n$ represents the current inner iteration (line \ref{algo: sle}). The reconstruction might be quite noisy, (as we show in Sec. \ref{sec: mock nis reco}), nonetheless, the results in the overall reconstruction can be improved by introducing a smoothing on $\psi^{(n)}$. Such smoothing is the result of recomputing $\psi^{(n)}$ with equation \eqref{eq: finite diff pot}, where for this case $\psi_i$ represents the smoothed potential at the $i$-th node, while $\psi_j$ comes from the solution of \eqref{eq: system of linear equations}. This step is optional but advised, as we justify in Sec. \ref{sec: mock nis reco} (line \ref{algo: pot_sm}). 

Then, from $\psi^{(n)}$ (smoothed or not) we compute the convergence $\kappa^{(n)}$ on the grid (line \ref{algo: kappa}) by using finite differences, such that, when the condition
\begin{equation}\label{eq: inner convergence}
\Big|\kappa^{(n)}_{j}-\kappa^{(n-1)}_{j}\Big|\leq \text{tolerance}
\end{equation}
is satisfied for every node ($j=1,\dots,N$), (line \ref{algo: converge}), the inner level finishes its task (line \ref{algo: inner end}). In \eqref{eq: inner convergence} $\kappa^{(n-1)}$ represents the convergence from the previous inner iteration. 

When the inner level finishes, we compute either the type 1 or type 2 penalty function $\chi^2_s$ depending on the initial conditions (line \ref{algo: penalty}). We have seen that optimal results are obtained when the reconstruction reaches $\chi^2_s\sim 2N_{\text{img}}$, with $\text{N}_{\text{img}}$ being the total number of multiple images being used. This condition is intended to avoid an overfit, without compromising the quality of the reconstruction.   With that in mind, we consider that once the condition $\chi^2_s\leq q (2N_{\text{img}})$ (for $q>1$) is satisfied (line \ref{algo: break}), the outer level as well as the reconstruction finish (line \ref{algo: outer end}). Here we have taken $q=1.4$, for which the reconstruction tends to satisfy $(q-1)(2N_{\text{img}})\leq\chi^2_s\leq q (2N_{\text{img}})$.

If condition $\chi^2_s\leq q(2N_{\text{img}})$ is not met, a new outer iteration is required (line \ref{algo: outer new}). Thus, the grid refinement is performed (line \ref{algo: refinement}) (more details in Sec. \ref{sec: grid refinement}). Then, from the deflection potential result of the last inner iteration for the current outer iteration, we compute $\psi^{(m)}$, $\kappa^{(m)}$, $\gamma_1^{(m)}$ and $\gamma_2^{(m)}$ on this new grid (line \ref{algo: new value}) (more details in Sec. \ref{sec: finite difference}). They are required to initialize the next outer iteration (more details in Sec. \ref{sec: penalty function}), taking the place of $\psi^{(m-1)}$, $\kappa^{(m-1)}$, $\gamma_1^{(m-1)}$ and $\gamma_2^{(m-1)}$, or $\psi^{(0)}$, $\kappa^{(0)}$, $\gamma_1^{(0)}$ and $\gamma_2^{(0)}$ if we are running the first outer iteration. Here, $m$ stands for the current outer iteration.

Finally, if no further outer iterations are required, as return we have the final deflection potential, which corresponds to the last $\psi^{(n)}$ that we know (line \ref{algo: return}).

Due to the nature of the method, negative densities may appear. This happens mainly for blind reconstruction (i.e. $\psi^{(0)}_{j}=0$ at every node), as well as in situations with little observational constraints. The negative values tend to appear towards the outskirts of the distribution. The presence of negative densities is also heavily related to the initial conditions. For instance, small values of $\eta_{\kappa}$ and $\eta_{\gamma}$ produce an overfit of the intrinsic noise in the reconstruction. Likewise, a too dense initial grid restrains the capability of the method to adapt to the data.  One way to solve this problem (when it occurs), is to use as an initial guess a distribution other than the blind one, which, along with the smoothing reduce the noise in the reconstruction, particularly towards the outskirts.  

\subsection{Finite difference approach}\label{sec: finite difference}
In this regard, we follow the generalized finite difference approach (GFD) (e.g. \cite{benito2001influence, gavete2003improvements, benito2003h}), with the difference that we use it not only for finding the derivatives on a grid from the deflection potential defined on such a grid, but also for interpolating the deflection potential and its derivatives at a given angular position $\boldsymbol{\theta}_0$, in case it is not part of the grid.

\begin{figure}
\centering
\includegraphics[scale = 0.55]{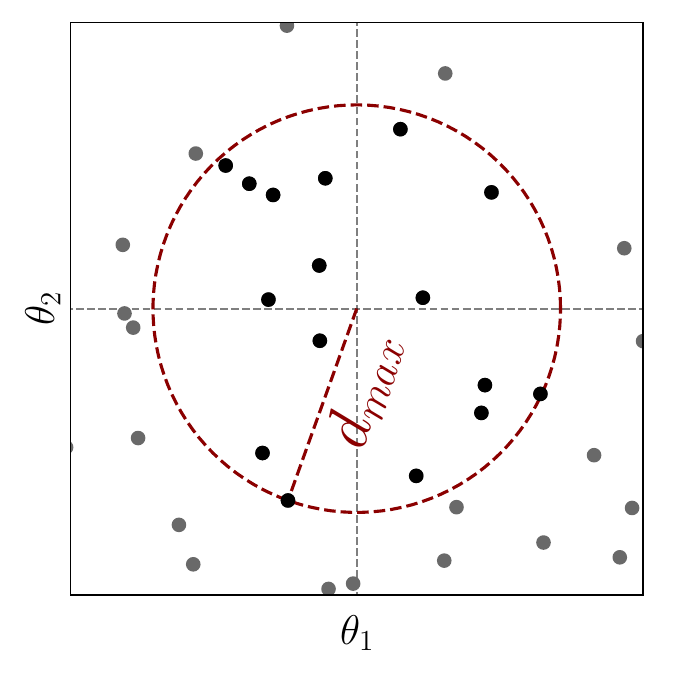}
\caption{In this figure the black dots represent the NN to $\boldsymbol{\theta}_0$, where the given dashed circle is centered at $\boldsymbol{\theta}_0$. The gray dots correspond to nodes that are not being  considered in the finite difference scheme.}
\label{fig: diff_near}
\end{figure}

Let us consider the $Q$ nearest nodes (NN) (nearest neighbors) to $\boldsymbol{\theta}_0$, as it is depicted in Fig. \ref{fig: diff_near}. On each of such NN we apply the Taylor series expansion to is corresponding deflection potential around $\boldsymbol{\theta}_0$ up to second order derivatives. Therefore, for the $i$-th NN we get
\begin{equation}\label{ew: taylor expand}
\psi_i = \psi_0 + h_i\dfrac{\partial \psi_0}{\partial\theta_1} + k_i\dfrac{\partial \psi_0}{\partial\theta_2} +\dfrac{h_i^2}{2}\dfrac{\partial^2 \psi_0}{\partial\theta_1^2} +h_ik_i\dfrac{\partial^2 \psi_0}{\partial\theta_1\partial\theta_2} +\dfrac{k_i^2}{2}\dfrac{\partial^2 \psi_0}{\partial\theta_1^2},
\end{equation}
with $h_i=\theta_{i,1}-\theta_{0,1}$, and $k_i=\theta_{i,2}-\theta_{0,2}$. In case for example flexion is being considered in the reconstruction, higher order derivatives have to be taken in the expansion \eqref{ew: taylor expand}. Here $X_1 = \psi_0$, $X_2 = \partial \psi_0/\partial\theta_1$, $X_3 = \partial \psi_0/\partial\theta_2$, $X_4 = \partial^2 \psi_0/\partial\theta_1^2$, $X_5 = \partial^2 \psi_0/\partial\theta_1\partial\theta_2$, and $X_ 6= \partial^2 \psi_0/\partial\theta_2^2$ are  the unknowns, which, following GFD, can be obtained by minimizing the penalty function
\begin{equation}\label{eq: GFD}
\chi^2_{GFD} = \sum_{i=1}^{Q}\Big(\Psi_i W_i\Big)^2,
\end{equation}
with $W_i=W(h_i,k_i)$ being a weighting function, and
\begin{equation}
\Psi_i = -\psi_i + X_1 + h_iX_2 + k_iX_3 +\dfrac{h_i^2}{2}X_4+h_ik_iX_5+\dfrac{k_i^2}{2}X_6. 
\end{equation}

Therefore, when minimizing \eqref{eq: GFD} with respect to the $j$-th unknown we get
\begin{equation}\label{eq: GFD minimization}
\dfrac{\partial\chi^2_{GFD}}{\partial X_j } = 2\sum_{i=1}^{Q}\Psi_iW_i^2\dfrac{\partial \Psi_i}{\partial X_j}=0, 
\end{equation}
which leads to a system of linear equations of the form $A\boldsymbol{X}=\boldsymbol{b}$, with
\begin{equation}\label{eq: GFD system}
A=\displaystyle\sum_{i=1}^Q W_i^2\left(\begin{array}{c c c c c c}
1 & h_i & k_i & \dfrac{h_i^2}{2} & h_ik_i & \dfrac{k_i^2}{2}\\[2mm]
& h_i^2 & h_ik_i & \dfrac{h_i^3}{2}& h_i^2k_i & \dfrac{h_ik_i^2}{2}\\[2mm]
&  & k_i^2 & \dfrac{h_i^2k_i}{2}& h_ik_i^2 & \dfrac{k_i^3}{2}\\[2mm]
& \mathrm{SYM} &   & \dfrac{h_i^4}{4}& \dfrac{h_i^3k_i}{2} & \dfrac{h_i^2k_i^2}{4}\\[2mm]
&  &   &  & h_i^2k_i^2 & \dfrac{h_ik_i^3}{2}\\[2mm]
&  &   &  &  & \dfrac{k_i^4}{4}
\end{array}\right),
\end{equation}
and
\begin{equation}
\boldsymbol{b}^{T}=\sum_{i=1}^Q W_i^2\psi_i\left(1,\, h_i,\, k_i,\, h_i^2/2,\, h_ik_i,\, k_i^2/2\right),
\end{equation}
while the elements of $\boldsymbol{X}$ were defined above. We find $A^{-1}$ numerically, such that
\begin{equation}\label{eq: GFD solution}
X_j=\sum_{i=1}^{Q}\mathcal{X}_{ji}\psi_i,    
\end{equation}
where the weight of the $i$-th NN for the $j$-th unknown is given by
\begin{equation} \label{eq: GFD weight}
\mathcal{X}_{ji}=W_i^2\Bigg(A^{-1}_{j1}+A^{-1}_{j2}h_i+A^{-1}_{j3}k_i+A^{-1}_{j4}\dfrac{h_i^2}{2}+A^{-1}_{j5}h_ik_i+A^{-1}_{j6}\dfrac{k_i^2}{2}\Bigg).
\end{equation}

From \eqref{eq: GFD solution} it is clear that only the weights of the $Q$ NN are (in principle) different from zero. Here, from  \eqref{eq: GFD weight} it becomes straightforward to compute the weights required in equations \eqref{eq: finite diff pot} - \eqref{eq: finite diff alp}.

Now, we have found from mock reconstructions that for $Q\sim 9$ the reconstruction still works, but for less NN the reconstruction tends to fail, since \eqref{eq: GFD system} becomes singular quickly. Under this condition, if the reconstruction converges, most likely the output will not be reliable. On the other hand, if $Q$ is beyond $Q\sim 36$ the reconstruction flattens. We have seen that $16\leq Q\leq 36$ is good enough for the reconstruction to work properly, with optimal results for $Q \sim 25$, since the output is neither irregular in excess nor flattened in excess.

In this paper we use a quartic spline as $W_i$, given by
\begin{equation}\label{eq: weighting}
W_i = 1-6\left(\dfrac{d_i}{dm}\right)^2+8\left(\dfrac{d_i}{dm}\right)^3-3\left(\dfrac{d_i}{dm}\right)^4,
\end{equation}
with $d_i=\sqrt{(\theta_{i,1}-\theta_{0,1})^2+(\theta_{i,2}-\theta_{0,2})^2}$, and $dm=nd_{max}$, where $d_{max}$ is the distance of the furthest NN from $\boldsymbol{\theta}_0$ (as it is shown in Fig. \ref{fig: diff_near}), and $n>1$. In this work we have taken $n=2$.

\subsection{Grid refinement}\label{sec: grid refinement}
\begin{figure*}
\centering
\includegraphics[scale = 0.6]{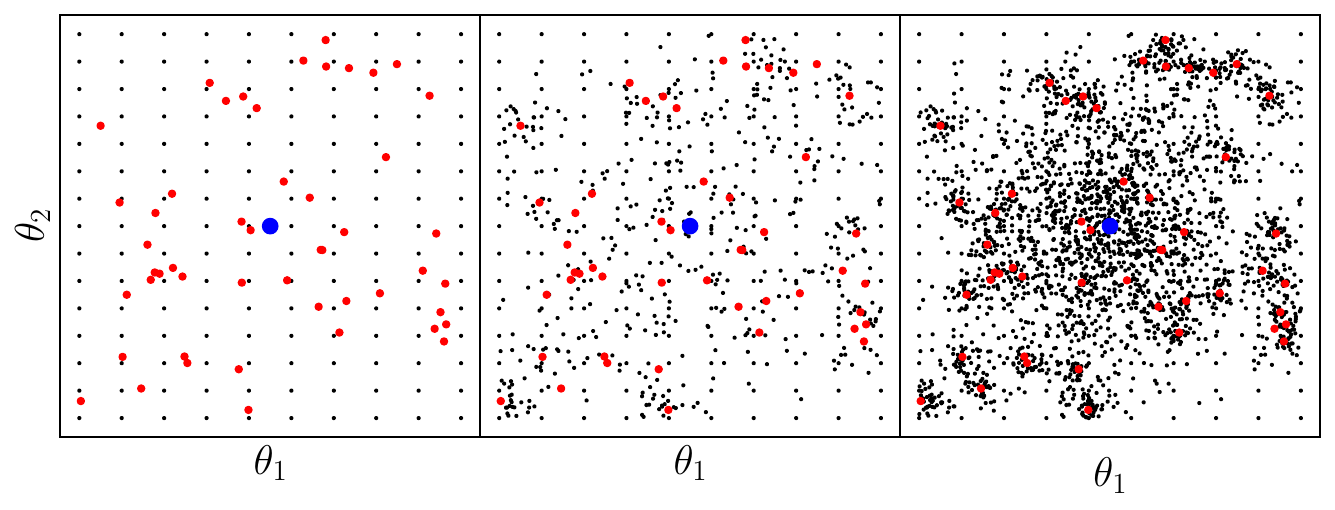}
\caption{In this figure the blue dot represents the position of the main deflector, the red dots represent the multiple images, and the black dots represent the nodes in the given grid. Here we have the initial grid (left panel), upon which the first refinement is applied at the end of the first outer level iteration (middle panel). The refinement continues until the reconstruction ends, giving us the final grid (right panel).}
\label{fig: refinement}
\end{figure*}

The reconstruction begins with a rectangular grid with $N_1 \times N_2$ nodes with respect to the rectangular coordinates $\theta_1$ and  $\theta_2$, defined over the region where the reconstruction takes place. 

For the refinement process, by means of a Gaussian distribution with standard deviation $\sigma_d$, we draw $N_d$ new nodes around each of the main deflectors when they are known. If that is not the case, by default, such new nodes are drawn with respect to the center of the region of reconstruction. We keep only those new nodes that lie within a radius $R_d>\sigma_d$ from the given main deflector. The aim of $R_d$ is to regulate the density of nodes, whereas weak lensing is dominant. With this distribution of nodes, the reconstruction is capable to provide a better description of those regions with high mass concentrations.

Additionally, around  each of the multiple images, we set a circular neighborhood of radius $r_d$ and determine the number of nodes within such neighborhood. If it is less than $Q$, by means of a uniform distribution, we add the number of nodes required to get $Q$ nodes. Otherwise, we do not add new nodes. For each outer iteration, $r_d$ is reduced by a factor $u$. This adaptive process provides a higher resolution, and thus a more accurate reproduction the multiply imaged systems. We apply this process only to the strong regime, since, in general terms the identification of the multiply imaged systems and their positions, as well as the penalty functions themselves,  are less noisy than those from the weak regime; and we want to avoid overfitting such noise as much as possible.

An example of the refinement process for a single deflector is depicted in Fig. \ref{fig: refinement}.

\subsection{Penalty functions}\label{sec: penalty function}
In this section we discuss how the different contributions to the penalty function \eqref{eq: penalty function} are defined, and how they contribute to \eqref{eq: system of linear equations}.

As we described before, the reconstruction method consists of two levels of iterations, where the system of linear equations is computed at the inner level. So that, when we compute such system of equations, those terms that mess up with the linearity with respect to the deflection potential are computed using the deflection potential from the previous inner iteration. In consequence, such terms can be considered as constants during the current inner iteration. If we are at the first inner iteration, we make use of   $\psi^{(m-1)}$, with $m\geq 1$ being the current outer iteration.

\subsubsection{Strong regime}\label{sec: penalty function strong}
The strong regime is constrained by using the multiply imaged systems available, as discussed in Sec. \ref{sec: lensig}. In particular, we focus on their angular position, which makes natural to define the penalty function for this regime as
\begin{equation}
\label{eq: penalty strong lens plane}
\chi^{2}_{s}:=\sum_{i=1}^{N_s}\Bigg(\sum_{n=1}^{N_i}\boldsymbol{p}_{in}^{T}\mathcal{S}^{-1}_{l,in}\boldsymbol{p}_{in}\Bigg),
\end{equation} 
where $N_s$ and $N_i$ are ,respectively, the number of multiply imaged systems that are being used, and the number of images the $i$-th system has. On the other hand, $\boldsymbol{p}_{in}:=\boldsymbol{\theta}_{in}-\boldsymbol{\theta}_{in}^{\prime}$ compares the $n$-th observed image position $\boldsymbol{\theta}_{in}^{\prime}$ from the $i$-th system to the corresponding position $\boldsymbol{\theta}_{in}$ computed from the reconstruction. The uncertainties in $\boldsymbol{\theta}_{in}^{\prime}$ at each component are $\sigma_{1,in}$ and $\sigma_{2,in}$, which are introduced trough 
\begin{equation}\label{eq: strong cov}
S_{l,in}=\left(\begin{array}{cc}
\sigma_{1,in}^{2} & 0\\
0 & \sigma_{2,in}^{2}
\end{array}\right).
\end{equation}

The problem with \eqref{eq: penalty strong lens plane} relies on the need to explicitly solve the lens equation, which makes difficult to explicitly write it in terms of $\psi$ on every node. For that reason, working on the source plane instead, turns out to be an advantage. In this direction, let us define two different penalty functions.

\paragraph*{\textbf{Type 1:}} Since any multiply imaged system is linked to a unique source, when we apply the lens equation \eqref{eq: lens equation} to such images, they have to take us to the same source with position $\boldsymbol{\beta}$. Hence, for the  $i$-th system we can compare the source position $\boldsymbol{\beta}_{in}$  predicted by the reconstruction for the $n$-th image to the source true position $\boldsymbol{\beta}^{\prime}_{i}$, however, since such true position is unknown, the average source position predicted by the reconstruction
\begin{equation}\label{eq: source position average}
\braket{\boldsymbol{\beta}_{i}}=\frac{1}{N_i}\sum_{n=1}^{N_i}\boldsymbol{\beta}_{in}
\end{equation}
is commonly used instead. Here $\boldsymbol{\beta}_{in}=\boldsymbol{\theta}^{\prime}_{in}-Z_{i}\boldsymbol{\alpha}_{in}$. This approach also constrains the multiple images to converge to the same source as needed.

One cannot simply work on the source plane and make use of the observed uncertainties, since the effect of the lens is important. As it is discussed in \cite[Part 2, Section 4.6]{schneider2006gravitational}, the magnification induced on the images by the mapping $\boldsymbol{\beta}\to\boldsymbol{\theta}$ also affects their position uncertainties. Therefore, as a correction to this matter, by applying the Taylor expansion to the lens equation \eqref{eq: lens equation} up to first order derivatives around $\braket{\boldsymbol{\beta}_i}$\footnote{We use $\braket{\boldsymbol{\beta}_i}$ since the true source position $\boldsymbol{\beta}_i$ is unknown.}, we get
\begin{equation}\label{eq: lens equation expansion}
\boldsymbol{p}_{in}=\mathcal{M}_{in}\left(\boldsymbol{\beta}_{in}-\braket{\boldsymbol{\beta}_i}\right),
\end{equation}
where
\begin{align}\label{eq: magnificacion matrix}
\nonumber
\mathcal{M}_{in} 
=& \mu_{in}\left(\begin{array}{cc}
1-Z_{i}\kappa_{in}+Z_{i}\gamma_{1,in} & Z_{i}\gamma_{2,in}\\
Z_{i}\gamma_{2,in} & 1-Z_{i}\kappa_{in}-Z_{i}\gamma_{1,in}
\end{array}\right) \\
=& \mu_{in}\left(\begin{array}{cc}
m_{1,in} & m_{2,in}\\
m_{3,in} & m_{4,in}
\end{array}\right).
\end{align}
is the magnification matrix at $\boldsymbol{\theta}_{in}^{\prime}$, and
\begin{equation}\label{eq: manification}
\mu_{in}=\dfrac{1}{\left(1-Z_i\kappa_{in}\right)^2-Z_i^2\left(\gamma_{1,in}^2+\gamma_{2,in}^2\right)}
\end{equation}
is magnification the of the corresponding image as well. It is important to take into account that the approximation \eqref{eq: lens equation expansion} holds as long as the predicted images are as close as possible to the observed ones.

It has been shown that only the magnification $\mu_{in}$ is enough to account for this correction (e.g. \cite{bradac2005reco2}). We stick to this for the reconstructions presented in this work. Nevertheless, due to the nature of this reconstruction method, considering only the magnification (at least for our implementation) can lead to a divergent solution; the method may become numerically unstable. For instance, considering $\mathcal{M}_{in}$ as a hole helps to control the impact of $\mu_{in}$ making the reconstruction more stable, with the cost of losing some accuracy around the mass peaks. Now, the implementation of either $\mu_{in}$ or $\mathcal{M}_{in}$  introduce nonlinearities on \eqref{eq: system of linear equations}, so that they are not explicitly written in terms of $\psi_j$. Instead, they are computed from the last $\psi$ that is known, and are manipulated as constant terms.

With respect to $\braket{\boldsymbol{\beta}_i}$, one can compute it in the same way as $\mu_{in}$ and $\mathcal{M}_{in}$ are computed (as it is done in e.g. \cite{bradac2005reco, bradac2009cosmictelescope}). However, at least for our implementation this approach is not effective, since when either $\mu_{in}$ or $\mathcal{M}_{in}$ is included the reconstruction shows the tendency to produce an overestimation of $\kappa$, close to some of the grid corners, and it cannot be effectively solved simply by considering a wider region for the reconstruction or changing the grid size. This problem is solved by explicitly writing the components of $\braket{\boldsymbol{\beta}_i}$ in terms of $\psi_j$ as
\begin{equation}
\braket{\beta_{q,i}}=\dfrac{1}{N_i}\sum_{p=1}^{N_i}\theta_{q,ip}^{\prime}-\dfrac{1}{N_i}\sum_{p=1}^{N_i}\Bigg(\sum_{j=1}^{N}\mathcal{D}_{q,ipj}\psi_{j}\Bigg),  
\end{equation}
from which we have that 
\begin{equation}
\braket{\theta_{q,i}} := \dfrac{1}{N_i}\sum_{p=1}^{N_i}\theta_{q,ip}^{\prime},
\end{equation}
and
\begin{equation}
\dfrac{\partial\braket{\beta_{q,i}}}{\partial\psi_k} := -\dfrac{Z_i}{N_i}\sum_{p=1}^{N_i}\mathcal{D}_{q,ipk}, 
\end{equation}
with $q=1,2$ representing each component. Under this approach, we also get an improvement in the  performance. For a tolerance of $10^{-3}$ we observed in the different reconstructions presented in this work a reduction in the number of inner iterations of at least ten times. Also, structure was clearly shown after the first or second inner iteration, and the condition \eqref{eq: inner convergence} was fulfilled within $4-6$ inner iterations. 

At this point, we have that our first penalty function is easily obtained by replacing  \eqref{eq: lens equation expansion} into \eqref{eq: penalty strong lens plane}. Having $\chi_s^{2}$, its contribution to \eqref{eq: system of linear equations} is given through
\begin{equation}
\label{eq: w strong 1}
\mathcal{W}_{kj}^{(s)}
:=\sum_{i=1}^{N_s}\sum_{n=1}^{N_i}\sum_{q=1}^{2}Z_iB_{q,ink}\Bigg(
\mathcal{D}_{q,inj}
-\dfrac{1}{N_i}\sum_{p=1}^{N_i}\mathcal{D}_{q,ipj}\Bigg),
\end{equation}
and
\begin{align}
\label{eq: strong V 1}
\nonumber
\mathcal{V}_{k}^{(s)}
=&\sum_{i=1}^{N_s}\sum_{n=1}^{N_i}\sum_{q=1}^{2}\Bigg[b_{q,in}\Bigg(Z_i\mathcal{D}_{q,ink}+\dfrac{\partial\braket{\beta_{q,i}}}{\partial \psi_k}\Bigg)\\&
- B_{q,ink}\braket{\theta_{1,i}}\Bigg],
\end{align}
where we have defined
\begin{align}
\nonumber
B_{1,ink} := & \mu_{in}^2\Bigg(Z_i\Delta m_{1,in}D_{1,ink} + Z_i\Delta m_{2,in}D_{2,ink} \\&
+\Delta m_1\dfrac{\partial\braket{\beta_{1,i}}}{\partial \psi_k}
+\Delta m_2\dfrac{\partial\braket{\beta_{2,i}}}{\partial \psi_k}\Bigg),
\end{align}
\begin{align}
\nonumber
B_{2,ink} := & \mu_{in}^2\Bigg(Z_i\Delta m_{2,in}D_{1,ink} + Z_i\Delta m_{3,in}D_{2,ink} \\&
+\Delta m_2\dfrac{\partial\braket{\beta_{1,i}}}{\partial \psi_k}
+\Delta m_3\dfrac{\partial\braket{\beta_{2,i}}}{\partial \psi_k}\Bigg),
\end{align}
\begin{equation}
b_{1,in} :=
\mu_{in}^2\Big(\Delta m_{1,in}\theta_{1,in} + \Delta m_{2,in}\theta_{2,in}\Big),
\end{equation}
and
\begin{equation}
b_{2,in} :=
\mu_{in}^2\Big(\Delta m_{2,in}\theta_{1,in} + \Delta m_{3,in}\theta_{2,in}\Big),
\end{equation}
with 
\begin{equation}
\Delta m_{1,in} := \Bigg(\frac{m_{1,in}^2}{\sigma_{1,in}^2} + \frac{m_{3,in}^2}{\sigma_{2,in}^2}\Bigg),
\end{equation}
\begin{equation}
\Delta m_{2,in} := \Bigg(\frac{m_{1,in}m_{2,in}}{\sigma_{1,in}^2} + \frac{m_{3,in}m_{4,in}}{\sigma_{2,in}^2}\Bigg),
\end{equation}
and
\begin{equation}
\Delta m_{3,in} := \Bigg(\frac{m_{2,in}^2}{\sigma_{1,in}^2} + \frac{m_{4,in}^2}{\sigma_{2,in}^2}\Bigg).
\end{equation}

\paragraph*{\textbf{Type 2:}} In this case, for each of the multiply imaged systems, we compare the possible pairs formed from the position of the source predicted for each of the corresponding multiple images. Therefore, the penalty function reads
\begin{equation}\label{eq: chi2 strong 3}
\chi^{2}_{s}:=\sum_{i=1}^{N_s}\Bigg(\sum_{n=1}^{N_{i}-1}\sum_{m=n+1}^{N_i}\boldsymbol{b}_{inm}^{T}\mathcal{S}^{-1}_{l,inm}\boldsymbol{b}_{inm}\Bigg),
\end{equation} 
where $\boldsymbol{b}_{inm}=\boldsymbol{\beta}_{in}-\boldsymbol{\beta}_{im}$, with $\boldsymbol{\beta}_{in}=\boldsymbol{\theta}^{\prime}_{in}-Z_i\boldsymbol{\alpha}_{in}$ and  $\boldsymbol{\beta}_{im}=\boldsymbol{\theta}^{\prime}_{im}-Z_i\boldsymbol{\alpha}_{im}$. In order to account for the correction discussed above, that is needed in order to be able to work at source plane, we have defined
\begin{equation}\label{eq: cov 3}
S_{l,inm}=\dfrac{1}{2}\left(\begin{array}{cc}
\dfrac{\sigma_{1,in}^{2}}{\mu_{in}^{2}} + \dfrac{\sigma_{1,im}^{2}}{ \mu_{im}^{2}}& 0\\
0 &\dfrac{\sigma_{2,in}^{2}} {\mu_{in}^{2}} + \dfrac{\sigma_{2,im}^{2}}{\mu_{im}^{2}}
\end{array}\right).
\end{equation}

 Here, the contribution of $\chi_s^{2}$ to \eqref{eq: system of linear equations} yields 
 \begin{equation}
\mathcal{W}^{(s)}_{kj}:=\sum_{i=1}^{N_s}\sum_{n=1}^{N_i-1}\sum_{m=n+1}^{N_i}\sum_{q=1}^{2}\Bigg(\dfrac{Z_i^2}{\sigma_{q,inm}^2}
\Delta\mathcal{D}_{q,inmk}\Delta\mathcal{D}_{q,inmj}\Bigg),
\end{equation}
and
\begin{equation}
\mathcal{V}^{(s)}_{k}:=\sum_{i=1}^{N_s}\sum_{n=1}^{N_i-1}\sum_{m=n+1}^{N_i}\sum_{q=1}^{2}\Bigg(\dfrac{Z_i}{\sigma_{q,inm}^2}
\Delta\mathcal{D}_{q,inmk}\Delta\theta_{q,inm}\Bigg),    
\end{equation}
where we have defined
\begin{equation}
\Delta \mathcal{D}_{q,inmp}:=\mathcal{D}_{q,inp} -\mathcal{D}_{q,imp}
\end{equation}
with $p=k,j$, and
\begin{equation}
\Delta \theta_{q,inm}:=\theta^{\prime}_{q,in} -\theta^{\prime}_{q,im}.
\end{equation}
 
\subsubsection{Weak regime}\label{sec: penalty function weak}

Following \cite{bradac2005reco}, we constrain the weak regime by means of sources apparent ellipticity, as discussed in Sec. \ref{sec: lensig}. Therefore, for this regime the penalty function is defined as
\begin{equation}
\label{eq: penalty function-weak}
\chi^2_w:=\sum_{i=1}^{N_w}\frac{|\epsilon_i-\braket{\epsilon_i}|^2}{\sigma^2_i},
\end{equation}
with $N_w$ being the number of background galaxies for which their ellipticity and redshift is known. Additionally \cite{schneider2006gravitational}
\begin{equation}
\label{eq: sigma-weak}
\sigma^2_i=\dfrac{1}{N_{p,i}}\left(1-min\left(|g_i|^2,|g_i|^{-2}\right)\right)^2\sigma_{\epsilon^s}^2+\sigma^2_{err},
\end{equation}
where $N_{p,i}$ corresponds to the number of averaged apparent ellipticities in \eqref{eq: ellipticiy average}. Such ellipticities are from sources with the same redshift, and located in a neighborhood within which the properties of the lens do not change significantly \cite{bartelmann2001weak, schneider2006gravitational}. Now, since in practice it is not easy to fulfill these conditions, the best choice is to take $N_{p,i}=1$, and so consider $\epsilon_i$ to be the best representative of its sample at its neighborhood \cite{bradac2005reco}. That is why in \eqref{eq: penalty function-weak} $\epsilon_i$ is compared to  $\braket{\epsilon_i}$. The quantities $\sigma_{\epsilon_s}$ and $\sigma_{err}$ in \eqref{eq: sigma-weak} correspond to the intrinsic and observational ellipticity standard deviations, respectively. In this paper we take $\sigma_{\epsilon_s}\sim 0.2-0.3$ and $\sigma_{err}=0.1$ \cite{bradac2005reco, cain2016flexion}.

An alternative approach to $\chi_{w}^2$ can be found in e.g. \cite{merten2009reco,merten2011reco2,merten2016freemesh}.

According to \eqref{eq: ellipticiy average}, the penalty function \eqref{eq: penalty function-weak} has two possible forms. To begin with, for $|g_i| \le 1$ \eqref{eq: penalty function-weak} turns into
\begin{align}
\label{eq: penalty weak<=}
\chi^2_w=
&\sum^{N_w}_{i=1}\frac{1}{\sigma_i^2}\Bigg|\epsilon_i-\frac{Z_i\gamma_i}{1-Z_i\kappa_i}\Bigg|^2=\sum^{N_w}_{i=1}\dfrac{\Big|\epsilon_i-Z_i\epsilon_i\kappa_i-Z_i\gamma_i\Big|^2}{\left(1-Z_i\kappa_i\right)^2\sigma_i^2},
\end{align}
whereas for $|g_i| > 1$ \eqref{eq: penalty function-weak} becomes
\begin{align}
\label{eq: penalty weak>}
\chi^2_w=
&\sum^{N_w}_{i=1}\dfrac{1}{\sigma^2}\Bigg|\epsilon_i-\frac{1-Z_i\kappa_i}{Z_i\gamma_i^{*}}\Bigg|^2=\sum^{N_w}_{i=1}\frac{\Big|Z_i\epsilon_i\gamma_i^{*}+Z_i\kappa_i-1\Big|^2}{Z_i^2|\gamma_i|^2\sigma_i^2}.
\end{align}

In practice, the observations mainly come from sources that satisfy $|g_i| < 1$. However, we have implemented both possibilities in \relensing, such that, for every inner iteration we evaluate $|g_i|$ from the last deflection potential that we have. With that, \relensing\ decides whether to use \eqref{eq: penalty weak<=} or \eqref{eq: penalty weak>} as the corresponding penalty function for the weak regime.  

For both versions of $\chi^2_{w}$ the coefficients $\mathcal{W}^{w}_{kj}$ and $\mathcal{V}^{w}_{k}$ can be written as
\begin{align}
\label{eq: penalty weak-derivative-B}
\nonumber
\mathcal{W}_{kj}^{(w)}
:=&\sum^{N_w}_{i=1}\Bigg[
A_{1,i}\Bigg(\mathcal{G}_{1,ik}\mathcal{K}_{ij}+\mathcal{K}_{ik}\mathcal{G}_{1,ij}\Bigg)\\\nonumber
&+A_{2,i}\Bigg(\mathcal{G}_{2,ik}\mathcal{K}_{ij}+{K}_{ik}\mathcal{G}_{2,ij}\Bigg)
+A_{3,i}\mathcal{G}_{1,ik}\mathcal{G}_{1,ij}\\
&+A_{4,i}\mathcal{G}_{2,ik}\mathcal{G}_{2,ij}+A_{5,i}\mathcal{K}_{ik}\mathcal{K}_{ij},
\Bigg],
\end{align}
and
\begin{align}
\mathcal{V}_{k}^{(w)}
:=&\sum^{N_w}_{i=1}\Bigg(a_{1,i}\mathcal{G}_{1,ik}+a_{2,i}\mathcal{G}_{2,ik}+a_{3,i}\mathcal{K}_{ik}\Bigg),
\end{align}
such that for $|g_i|\le 1$  we get
\begin{align}\nonumber
&A_{1,i}=\frac{Z_i^2}{\sigma_{\le,i}^2}\epsilon_{1,i},\quad A_{2,i}=\frac{Z_i^2}{\sigma_{\le,i}^2}\epsilon_{2,i},\quad
A_{3,i}=A_{4,i}=\frac{Z_i^2}{\sigma_{\le,i}},\\\nonumber&
A_{5,i}=\frac{Z_i^2}{\sigma_{\le,i}^2}|\epsilon_{i}|^2,\quad
a_{1,i}=\frac{Z_i}{\sigma_{\le,i}^2}\epsilon_{1,i},\quad a_{2,i}=\frac{Z_i}{\sigma_{\le,i}^2}\epsilon_{2,i},\quad\\& a_{3,i}=\frac{Z_i}{\sigma_{\le,i}^2}|\epsilon_{i}|^{2},
\end{align}
meanwhile for $|g_i|>1$ the coefficients are
\begin{align}\nonumber
&A_{1,i}=\frac{Z_i^2}{\sigma_{>,i}^2}\epsilon_{1,i},\quad
A_{2,i}=\frac{Z_i^2}{\sigma_{>,i}^2}\epsilon_{2,i},\quad
A_{3,i}=A_{4,i}=\frac{Z_i^2}{\sigma_{>,i}^2}|\epsilon_{i}|^2,\quad \\\nonumber&
A_{5,i}=\frac{Z_i^2}{\sigma_{>,i}^2},\quad 
a_{1,i}=\frac{Z_i}{\sigma_{>,i}^2}\epsilon_{1,i},\quad a_{2,i}=\frac{Z_i}{\sigma_{>,i}^2}\epsilon_{2,i},\\& a_{3,i}=\frac{Z_i}{\sigma_{>,i}^2},
\end{align}
with $\sigma^2_{\le,i} :=  (1-Z_i\kappa_i)^2\sigma^2_{i}$ and $\sigma_{>,i}^2 := Z_i^2|\gamma_i|^2\sigma_i^2$. Here, both $\sigma^2_{\le,i}$ and $\sigma^2_{>,i}$ are computed from the last $psi$ that is known, so that they are manipulated as constant terms.

\subsubsection{Regularization terms}\label{sec: penalty function regularization}

This sort of grid based reconstruction methods require a term which provides stability along the reconstruction. Such term helps to deal with numerical noise, which  otherwise will most likely lead to divergent solutions for \eqref{eq: system of linear equations}. 

Different solutions have been proposed to overcome these difficulties, either as a regularization term (e.g \cite{seitz1998entropy,bradac2005reco, bradac2005reco2, merten2016freemesh}), or by introducing the signal to noise ratio (e.g \cite{deb2008reconstruction}). Here, as we have stated before,  we have chosen to implement the former approach, particularly by using the regularization terms
\begin{equation}\label{eq: reg_k}
	\chi^2_{\kappa(R)} = \eta_{\kappa}\sum_{j=1}^{N}\Big(\kappa_j-\kappa_{j}^{(m-1)}\Big)^2,
\end{equation}
and
\begin{equation}
\chi^2_{\gamma(R)} 
= \eta_{\gamma}\sum_{j=1}^{N}\sum_{q=1}^2\Big(\gamma_{q,j}-\gamma_{q,j}^{(m-1)}\Big)^2,
\label{eq: reg_sh}
\end{equation}
with $\eta_{\kappa}$ and $\eta_{\gamma}$ being positive defined constants. Such constants or regularization weights, control how smooth the reconstruction turns. The term \eqref{eq: reg_k} acts mainly within the cluster inner region dominated by the strong regime, while the term \eqref{eq: reg_sh} helps to control the reconstruction outside the inner region. 

On the other hand, if we are in the $m$-th outer iteration, the terms $\kappa_{j}^{(m-1)}$, $\gamma_{1,j}^{(m-1)}$, and $\gamma_{2,j}^{(m-1)}$  in \eqref{eq: reg_k} and \eqref{eq: reg_sh}, respectively, are computed after the grid refinement at the end of the previous outer iteration, or they come from the initial guess ($m=1$).

Therefore, from \eqref{eq: reg_k} we get the coefficients
\begin{equation}
\mathcal{W}_{kj}^{(\kappa)}:=\eta_{\kappa}\sum_{i=1}^{N}\mathcal{K}_{ik}\mathcal{K}_{ij} 
\end{equation}
and
\begin{equation}
\mathcal{V}_k^{(\kappa)}:=\eta_{\kappa}\sum_{i=1}^{N}\kappa_i^{(m-1)}\mathcal{K}_{ik},
\end{equation}
meanwhile for \eqref{eq: reg_sh} we have
\begin{equation}
\mathcal{W}_{kj}^{(\gamma)}:=\eta_{\gamma}\sum_{i=1}^{N}\sum_{q=1}^{2}\mathcal{G}_{q,ik}\mathcal{G}_{q,ij},
\end{equation}
and\begin{equation}
\mathcal{V}_k^{(\gamma)}:=\eta_{\gamma}\sum_{i=1}^{N}\sum_{q=1}^{2}\gamma_{q,i}^{(m-1)}\mathcal{G}_{q,ik}.
\end{equation}

\section{Mock distributions: simple models}\label{sec: simple models}
For testing the performance of \relensing\ we apply the reconstruction to two mock lenses, modeled by means of Non-singular Isothermal Spheres (NIS for short). Below, we describe how the mock catalogues are created, and discuss the outcomes from their corresponding reconstructions. 

Hereinafter, for comparing the reconstructions to the true distribution, we use the relative difference $\Delta_{r}X = (X_{reco}-X_{true})/X_{true}$, defined for any quantity $X$ of interest.

\subsection{Catalogues}\label{sec: mock catalogues}
For each model, the catalogues involved are drawn under the same conditions.

For the strong regime, we consider $40$ sources randomly selected with  redshift $z_s>z_l$, and for which more than one image is obtained after solving the lens equation \eqref{eq: lens equation}. Whereas for the weak regime we consider $1000$ background sources with redshift $z_w>z_l$, whose ellipticities catalogue is constructed in the same direction as in \cite{bradac2005reco,cain2016flexion}. Thus,  the intrinsic ellipticities are drawn from a Gaussian distribution with $\sigma_{\epsilon^{s}}=0.2$, then, to each component of the observed ellipticities (which are computed from \eqref{eq: ellipticiy}) an error is added. Such an error is drawn from a Gaussian distribution with $\sigma_{err}=0.1$. Their redshifts are drawn from a Gamma distribution with shape and rate parameters $\alpha=3$ and $\beta =3/2$, respectively. For each redshift, the corresponding added uncertainty is drawn from a Gaussian distribution with $\sigma_{z,i}=0.05(1+z_i)$ (with $i=1,\dots,N_w$). We keep those sources with only one image after solving the lens equation.

\subsection{NIS and 2NIS}
The deflection potential for a NIS is defined as
\begin{equation}\label{eq: potential nis}
\psi(\boldsymbol{\theta}) = R_0\sqrt{|\Delta\boldsymbol{\theta}|^2+\theta_0^2}\quad\text{with}\quad R_0:=\frac{4\pi \sigma^2 D(z_l,z)}{c^2 D(z)},
\end{equation}
where $\Delta\boldsymbol{\theta}:=\boldsymbol{\theta}-\boldsymbol{\theta}_{c}$. Such deflection potential depends on the angular diameter distance between the lens and the given source ($D(z_l,z)$), and also between the observer and the source ($D(z)$). As well, it depends on the velocity dispersion $\sigma$, the speed of light $c$, the distribution core $\theta_0$, and distribution center $\boldsymbol{\theta}_c$.

The first model we consider is a NIS with redshift $z_l=0.4$, with center at $\boldsymbol{\theta}_{c}=(0, 0)\,arcmin$. It is characterized by the parameters $\theta_{0}=0.4\,arcmin$ and $\sigma =1500\,km/s$. The corresponding convergence map for a source with redshift $z_r=9$ is depicted in Fig. \ref{fig: nis convergence} (True). 

The second model consists of two identical NIS with redshift $z_l=0.4$, whose center is at $\boldsymbol{\theta}_{c1}=(-0.2,-0.4)\,arcmin$,  and $\boldsymbol{\theta}_{c2}=(0.2,0.4)\,arcmin$, respectively. The parameters describing each of those NIS are $\theta_{0}=0.2\,arcmin$ and $\sigma =1000\,km/s$, such that the resulting convergence map for a source with redshift $z_r=9$ is shown in Fig. \ref{fig: 2nis convergence} (True). We refer to this model as 2NIS.

\subsection{Reconstruction}\label{sec: mock nis reco}
The reconstructions are carried out considering  $\eta_{\kappa}=\eta_{\gamma} = 200$ irrespectively of the initial guess $\psi^{(0)}$. Here, for the NIS we only consider blind reconstructions, and then we explore the effect of the smoothing on such reconstructions. Whereas, for the 2NIS we compare blind reconstructions against those for which $\psi^{(0)}$ is given by a Non-singular Isothermal Ellipsoid (NIE for short), as we discuss later.

Now, for the initial grid, when only the strong regime is being considered (\textbf{s}), we use $20\times 20$ nodes over a region of $3\times3 \,arcmin^2$, meanwhile, when both regimes are being considered (\textbf{s+w}), we take $22\times 22$ nodes over a region of $3.2\times3.2 \,arcmin^2$. For the last scenario, the extra nodes and grid size help to avoid overestimations of the different quantities of interest on the boundary (particularly $\kappa$), which arise when some observations are close to such boundary. 

With respect to the refinement process, the NIS shows only one deflector, so that we add $N_d=600$ nodes for each refinement. On the other hand, the 2NIS consists of two deflectors, so that we add $N_d=300$ nodes for each deflector.  In both cases, those new nodes are distributed considering $\sigma_d = 1\,arcmin$, and $R_d = 2\,arcmin$. Also, we take $r_d = 0.2\,arcmin$ for the first  refinement process, which is reduced by a factor $u=0.8$ at each of the subsequent refinements, except for the last one for which the refinement is not applied. As we discussed in Sec. \ref{sec: finite difference}, we set $Q=25$ nodes to perform the finite differences. 

\subsubsection{Convergence} 
\begin{figure*}
\centering
\includegraphics[scale=0.5]{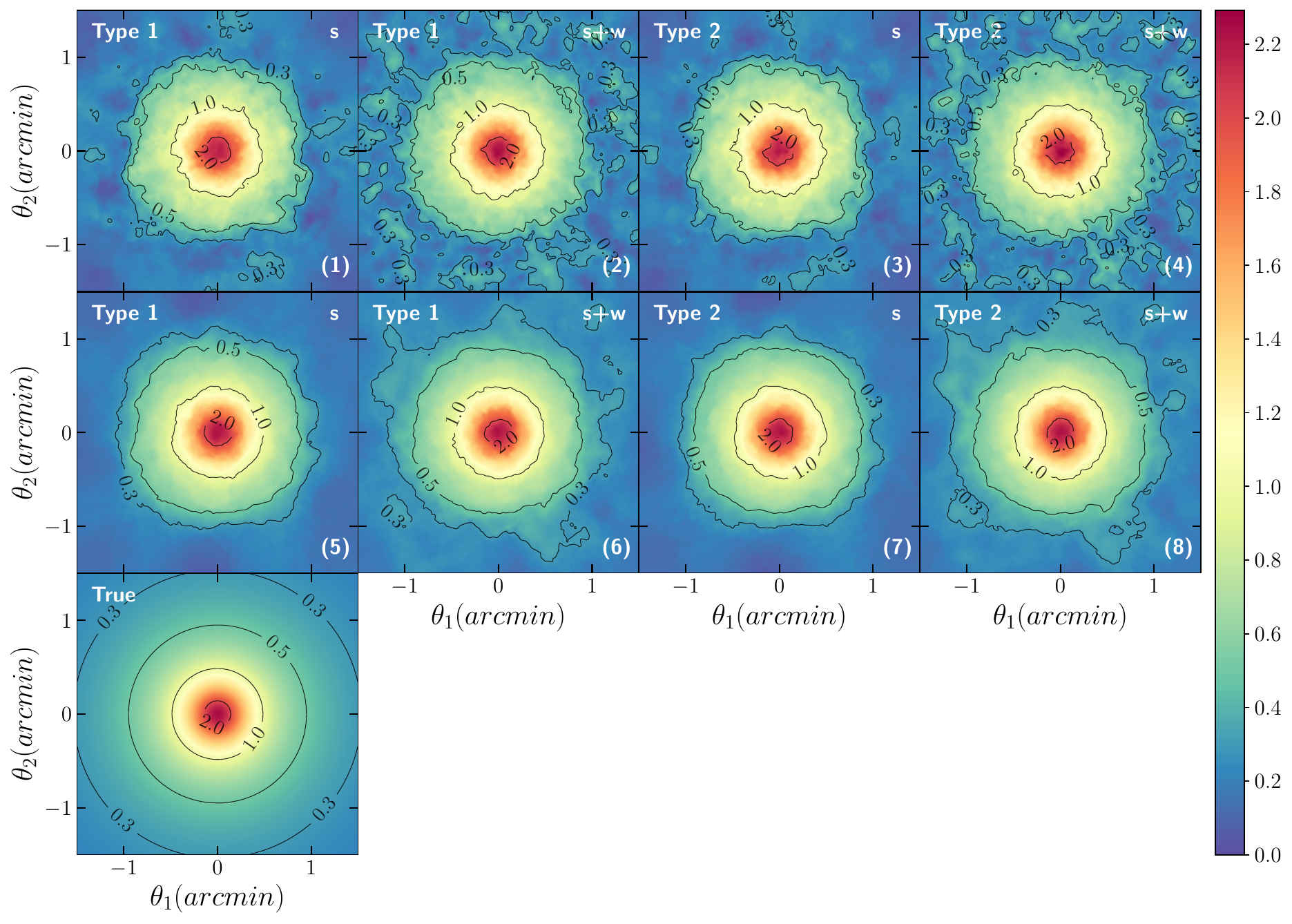}
\caption{Convergence maps ($\kappa$) for the NIS, given for a source with redshift $z_r=9$. The upper row shows the blind reconstructions without considering the smoothing, while the middle row shows the blind reconstructions considering the smoothing. The true map is shown in the lower panel.}
\label{fig: nis convergence}
\end{figure*}

In Fig. \ref{fig: nis convergence} we can see eight different reconstructed convergence maps ($\kappa$), along with the true mass for the NIS; which is the simplest of the distributions that we are considering in this work. This distribution posses spherical symmetry, for which we have only considered blind reconstructions here, since a parametric fitting to the data from the strong regime either to a NIS itself or to a NIE leads to the true lens (or at least a distribution close to it), and that is not what we want. One of the aims of this work is to show the power of the method to reveal important features of the lens, based upon little knowledge about it, which is closer to what one has to face with real data.

\begin{figure*}
\centering
\includegraphics[scale =0.5]{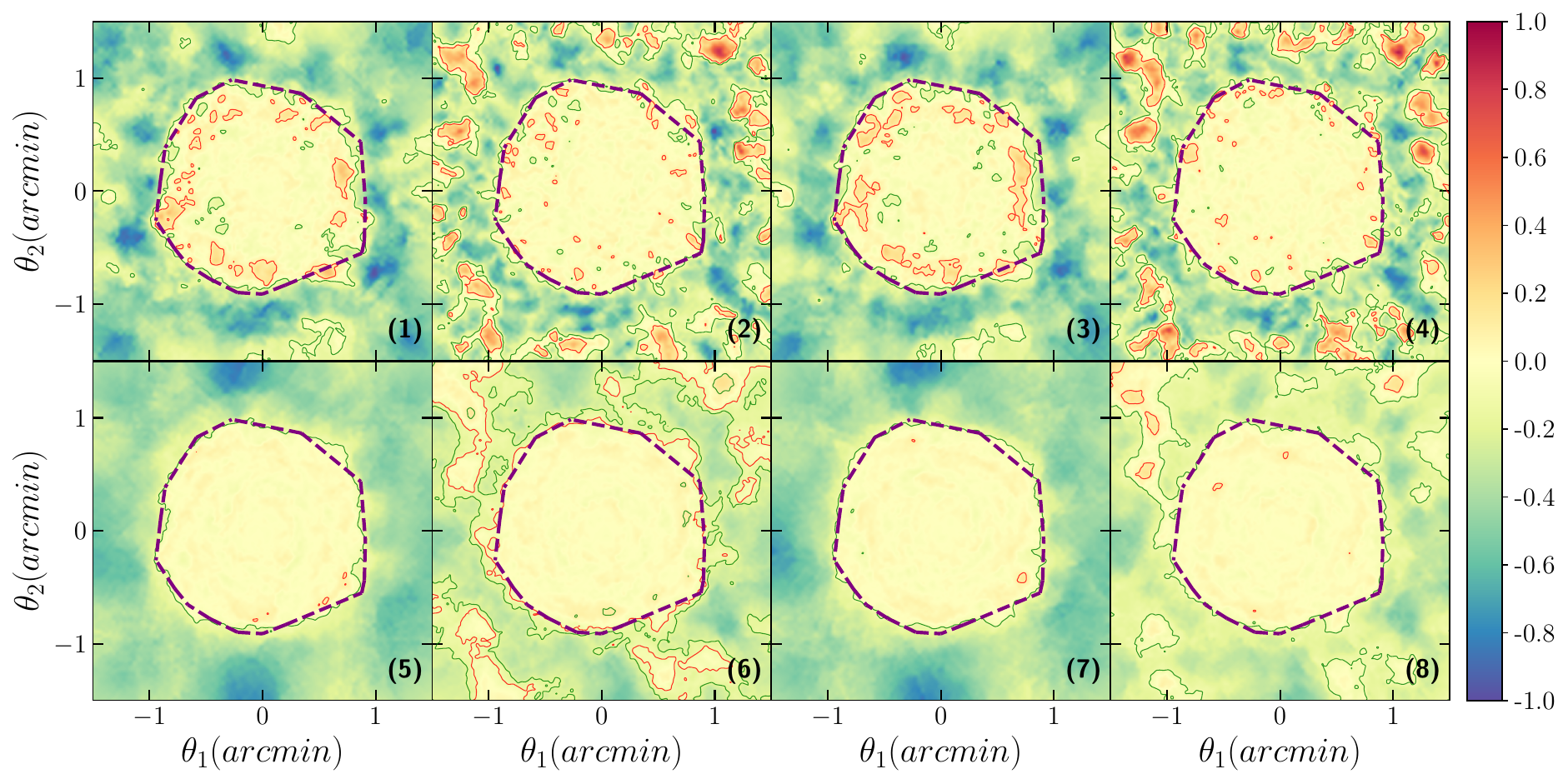}
\caption{Relative difference $\Delta_{r}\kappa$ between the reconstructed and true convergence maps for the NIS shown in Fig. \ref{fig: nis convergence}. Here, the red and green solid contours correspond to $\Delta_{r}\kappa=0.1$ and $\Delta_{r}\kappa=-0.1$, respectively. Meanwhile, the purple dashed contour delimits the region within which we have multiple images available.}
\label{fig: nis convergence comp}
\end{figure*}

Let us start with the effect of the smoothing. The Fig. \ref{fig: nis convergence} shows in the upper row the reconstructions for both penalty functions $\chi^{2}_{s}$ without considering the smoothing (reconstructions $(1)-(4)$), meanwhile, in the middle row the reconstructions were carried out under the very same conditions, but with considering the smoothing (reconstructions $(5)-(8)$); so that reconstruction $(1)$ is directly compared to reconstruction $(5)$, and so on. It is clear that the reconstructions reveal the central peak regardless of the smoothing, the penalty function used, or whether weak lensing is being considered. However, in Fig. \ref{fig: nis convergence comp} we can see that the inclusion of the smoothing provides a better reconstruction once we move towards the outskirts.

Comparing Type 1 and Type 2 reconstructions (carried out under the same conditions), we can see that there are no significant differences when it comes to the results given by both penalty functions. Instead, a significant difference appears when the smoothing is included. Without smoothing, the contours in $\kappa$ are noisier and less accurate, particularly outside the inner region. This is clearer when weak lensing constraints are being considered. Here, for the inner region we refer to the region that encloses all the multiple images known, as it is depicted in Fig. \ref{fig: nis convergence comp} (purple dashed contour). That behaviour stands irrespectively of the lens distribution or the input deflection potential $\psi^{(0)}$. This behavior is accentuated when the complexity of the lens increases. 

When only the strong regime is included, the reconstruction effectiveness is strongly restricted to the inner region, as we can see in Fig. \ref{fig: nis convergence comp}, where most of such region presents a relative difference within $|\Delta_{r}\kappa|<0.1$ ($10\%$). In particular, we can see an improvement at the inner region boundary when the smoothing is included. Outside the inner region, the reconstruction loses its effectiveness and there is an underestimation of $\kappa$, which is less accentuated when the smoothing is added, but still present. Once strong and weak regimes are combined, there is an overall enhancement in the reconstruction outside the inner region, but without smoothing the increase in $\kappa$ is less uniform, and we can still see marked peaks of both underestimations and overestimations of $\kappa$. 

Now, we put our attention on $M=M(\leq\theta)$, which represents the mass enclosed within an angular radius $\theta$ measured from $\boldsymbol{\theta}=(0,0)\, arcmin$. This applies to all the reconstructions considered in this work. Here,  Fig. \ref{fig: nis mass} shows $M$ for the eight reconstructions, given for $\,arcmin \leq \theta\leq 1.5\,arcmin$. The curves have been separated into Type 1 (upper panel) and Type 2 (lower panel), so that we can directly compare the effects of using the strong/weak regime as well as the smoothing. We can see that $M$ behaves similarly for Type 1 and Type 2 reconstructions. Again, we see that the best performance appears within the inner region, where we can see a relative difference within $|\Delta_{r}M|\leq 0.05$ ($5\%$).

The vertical lines in Fig \ref{fig: nis mass}, given for $\theta_{in}\approx 0.86\,arcmin$ and $\theta_{out}\approx 1.04\,arcmin$, correspond, respectively, to the inner and outer radius of the inner region. It is within such limits, particularly close to $\theta_{out}$,  that $M$ starts to deviate from the true curve. Beyond the inner region, the effectiveness decreases, and for the range within which the reconstruction is being considered, the relative difference gets up to $|\Delta_{r}M|\approx 0.16$ $(16\%)$ for those reconstructions where only strong lensing is being used; irrespectively of the smoothing. With the inclusion of the weak regime, we see an improvement in $M$ of $\sim 7\%$.

We see that the estimation of $M$ is not severely affected by the smoothing, since the main features of the lens are recovered in both cases. However, the smoothing provides a clear improvement in the mass profile, as well as in estimation of the magnifications maps, and therefore the critical curves; as we discuss later. Hence, from here on we consider the smoothing in our reconstructions. 


 \begin{figure*}
\centering
\includegraphics[scale =0.5]{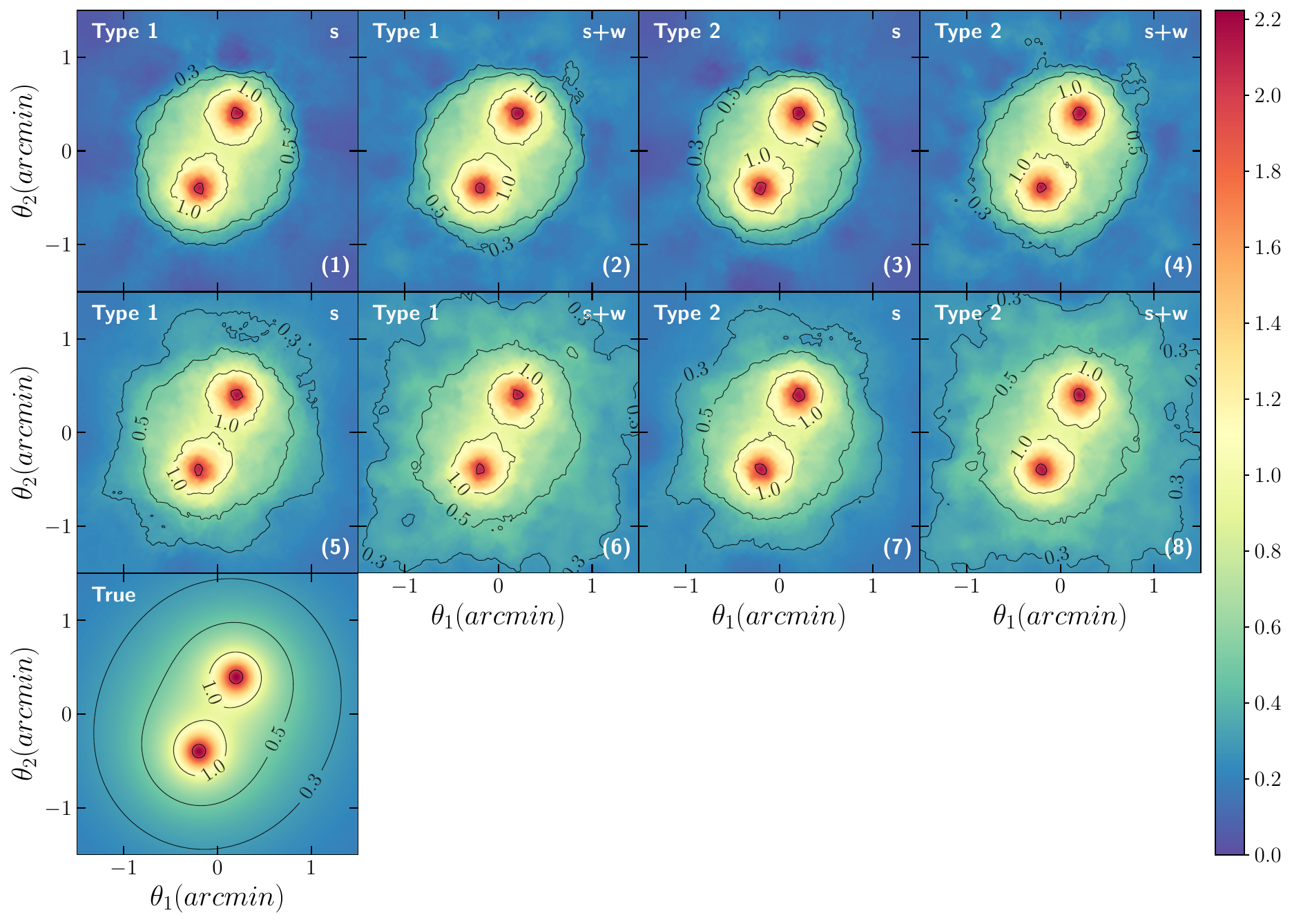}
\caption{Convergence maps ($\kappa$) for the 2NIS, given for a source with redshift $z_r=9$. The upper row shows the blind reconstructions, while the middle row shows the reconstructions with an input $\psi^{(0)}$ given by a NIE. For all reconstructions, the smoothing has been considered. The true map is shown in the lower panel.}
\label{fig: 2nis convergence}
\end{figure*}
\begin{figure*}
\centering
\includegraphics[scale =0.5]{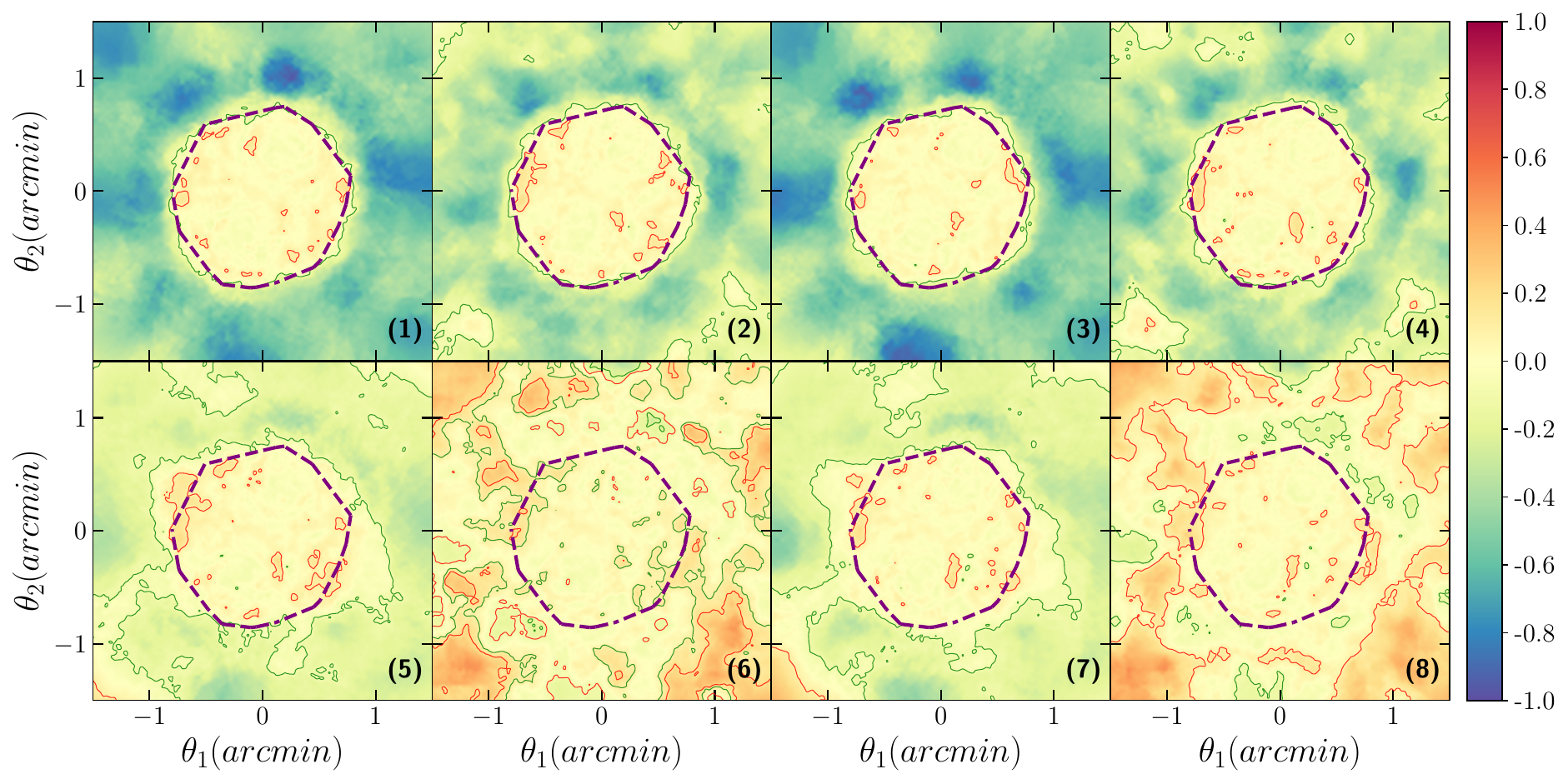}
\caption{Same as in Fig. \ref{fig: nis convergence comp} but for the 2NIS.}
\label{fig: 2nis convergence comp}
\end{figure*}

\begin{figure}
\centering
\includegraphics[scale =0.5]{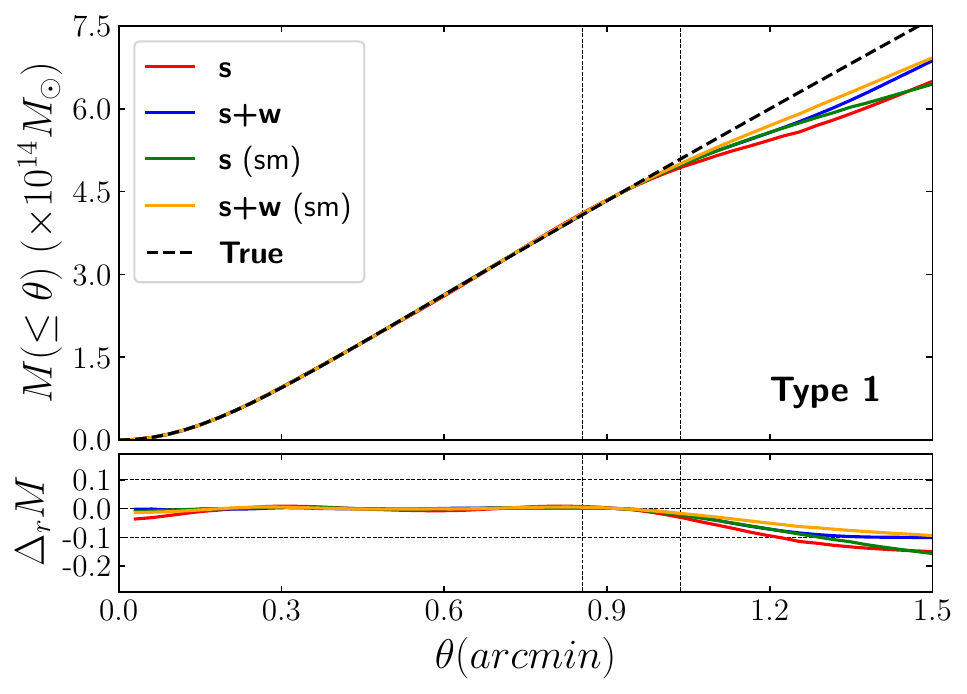}\\
\includegraphics[scale =0.5]{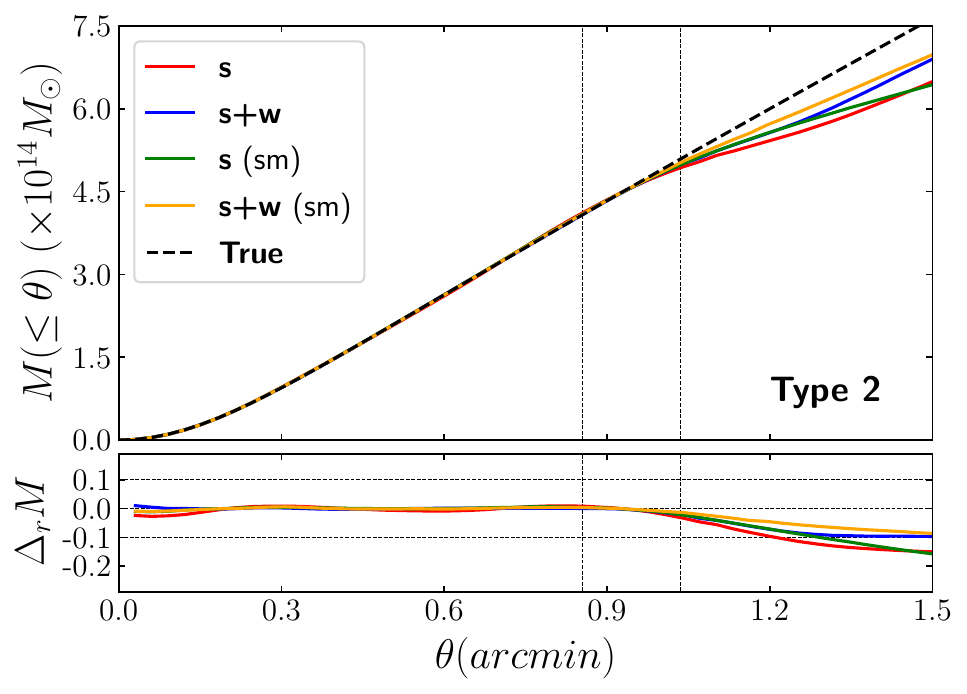}
 \caption{Mass for the NIS within an angular radius $\theta$, centered at $\boldsymbol{\theta} = (0,0)\,arcmin$ (upper panel in each figure), and the relative difference $\Delta_{r}M$ comparing the reconstructed and true mass (lower panel in each figure). The curves are separated depending on the penalty function $\chi^{2}_{s}$ used in the reconstruction, either \textbf{Type 1} or \textbf{Type 2}. The label \textbf{s} corresponds to reconstructions where only strong lensing was used, meanwhile the label \textbf{s+w} stands for reconstructions for which both, strong and weak lensing were used. Reconstructions with the label (sm) were carried out using the smoothing. The vertical dashed lines represent the inner and outer radius of the inner region (purple dashed contour in Fig. \ref{fig: nis convergence comp}). The horizontal dashed lines  enclose $|\Delta_{r}M|\leq 0.1$. Here, (sm) stands for those reconstructions for which the smoothing was applied.}
\label{fig: nis mass}
\end{figure}

\begin{figure}
\centering
\includegraphics[scale =0.5]{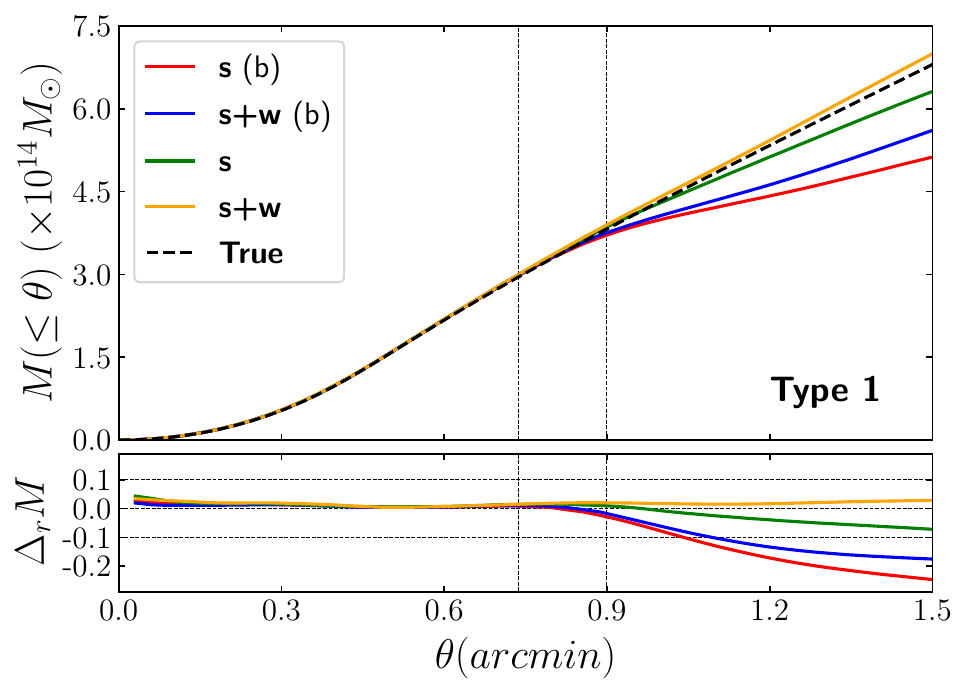}\\
\includegraphics[scale =0.5]{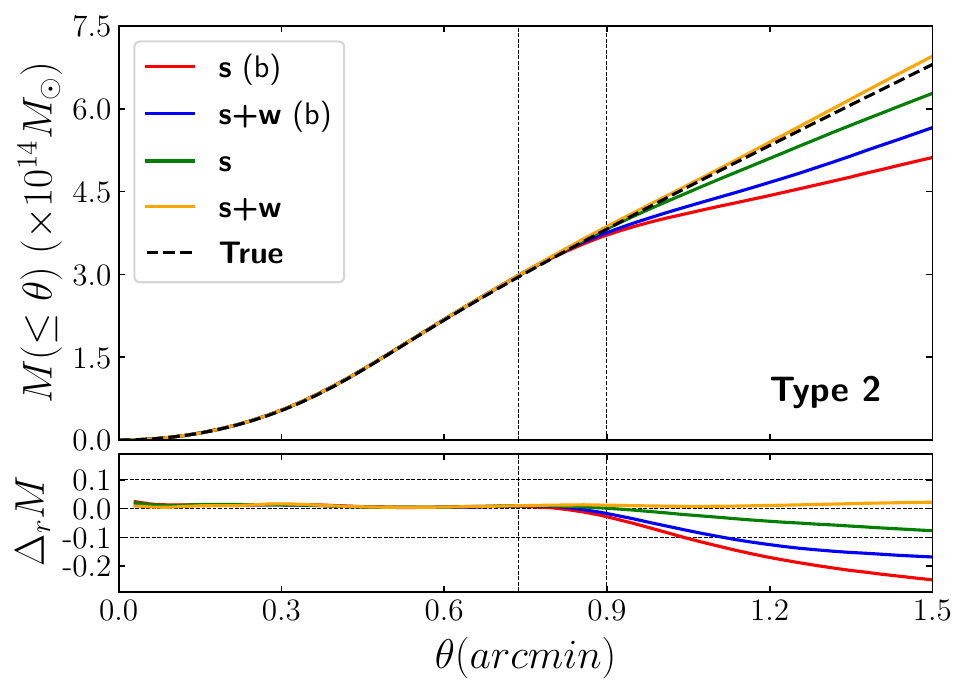}
\caption{Same as in Fig. \ref{fig: nis mass} but for the 2NIS, with the difference that here all reconstructions were carried out using the smoothing. Here, (b) stands for blind reconstruction.}
\label{fig: 2nis mass}
\end{figure}

For the 2NIS we focus on the comparison between blind reconstructions and those for which the input potential $\psi^{(0)}$ or initial guess contains some previous information about the lens. We want to keep this initial guess as simple as possible, so that we cannot interfere in excess, imposing many assumptions about the lens. For that reason, as it has been shown in \cite{bradac2005reco}, fitting a NIE to the multiply image systems provides a good enough estimation of $\psi^{(0)}$ to start working with. Depending on the complexity of the cluster, or its morphological properties, it may be convenient to use a combination of NIE as the initial guess. The deflection potential of such NIE is defined by making the transformation
\begin{align}\label{eq: nie transformation}
\nonumber
|\Delta\boldsymbol{\theta}|^2
\longrightarrow &
(1-e)\big(\Delta\theta_{1}\cos\varphi + \Delta\theta_{2}\sin\varphi\big)^2\\
&+(1+e)\big(-\Delta\theta_{1}\sin\varphi + \Delta\theta_{2}\cos\varphi\big)^2
\end{align}
into \eqref{eq: potential nis}, which accounts for the introduction of the ellipticity $e$ and the rotation $\varphi$ of the semi major axis with respect to the horizontal, and measured counterclockwise. For the fitting we have constrained $0\leq e\leq 0.25$, so the profile does not get distorted into a peanut shaped one, as it is discussed in e.g. \cite{pseudoellipticalnfw2002}. Here we have let $\theta_0=0.2\,arcmin$ fixed. Therefore, for the 2NIS, besides the blind reconstruction we consider a NIE as the initial guess.

In Fig. \ref{fig: 2nis convergence} we can see the eight reconstructions taken into account. The upper row shows the blind reconstructions (reconstructions $(1)-(4)$), whereas the middle row shows the reconstruction carried out under the same conditions except that the initial guess is given by the NIE (reconstructions $(5)-(8)$). The lower panel shows the true profile.

The 2NIS is characterized by two prominent peaks or main deflectors, which are effectively recovered in all reconstructions. Therefore, the initial guess is not relevant in this matter. In terms of the blind reconstructions, we can see that they struggle to recover the distribution beyond the inner region. It is expected for reconstructions $(1)$ and $(3)$, which only use strong lensing, as we saw for the NIS. In contrast, reconstructions $(2)$ and $(4)$, which also use weak lensing show an improvement in $\Delta_{r}\kappa$ of about $10\%-20\%$ towards the outskirts, as it is depicted in Fig. \ref{fig: 2nis convergence comp} (upper row). Within most of the inner region  $\kappa$ shows a relative difference $|\Delta_{r}|\kappa<0.1$, with few peaks that fall into $0.1 \,(10\%)\leq|\Delta_r\kappa|<0.2\,(20\%)$.

At this point, it is clear that weak lensing strengthens the reconstructions outside the inner region, so that the smoothing does not have a negative impact on such regime. We focus now on how the reconstruction behaves once one goes beyond blind reconstructions. For reconstructions $(5)-(8)$ we can see a considerable improvement with respect to their blind counterpart. Comparing reconstructions $(1)$ and $(5)$, as well as reconstructions $(3)$ and $(7)$ (which only use strong lensing), we find a closer approach to the true profiles, as it can be seen in the contour curves in $\kappa$ and also in the $\Delta_{r}\kappa$ maps in Fig. \ref{fig: 2nis convergence comp}. In addition, comparing reconstructions $(2)$ and $(6)$, as well as reconstructions $(4)$ and $(8)$ (which use strong and weak lensing), we can see that outside the inner region they go from an underestimation to an overestimation of $\kappa$, notwithstanding, the region where $|\Delta_{r}\kappa|<0.1$ is wider, and the overestimation lies beyond $|\Delta_{r}\kappa|>0.2$, mostly close to the boundaries, where the constraints are weaker. In either case, within the inner region, there are no significant changes in the reconstruction. Likewise, Type 1 and Type 2 reconstructions behave equally well. 

\begin{figure*}
\centering
\includegraphics[scale=0.52]{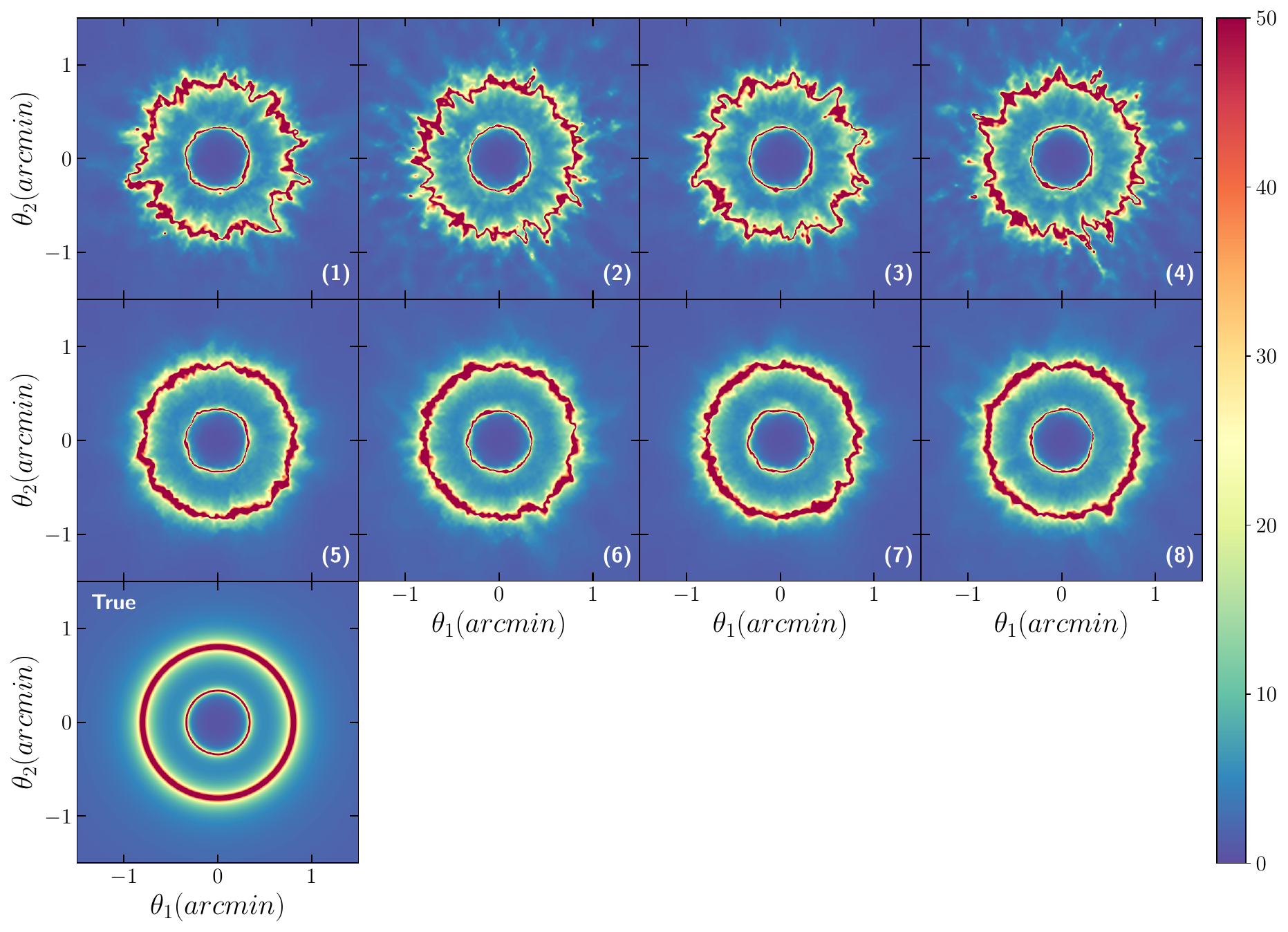}
\caption{Magnification maps ($|\mu|$) for the NIS, given for a source with redshift $z_r=9$. They correspond to the reconstructions shown in Fig. \ref{fig: nis convergence}. The true map is shown in the lower panel.}
\label{fig: nis magnification}
\end{figure*}
\begin{figure*}
\centering
\includegraphics[scale =0.52]{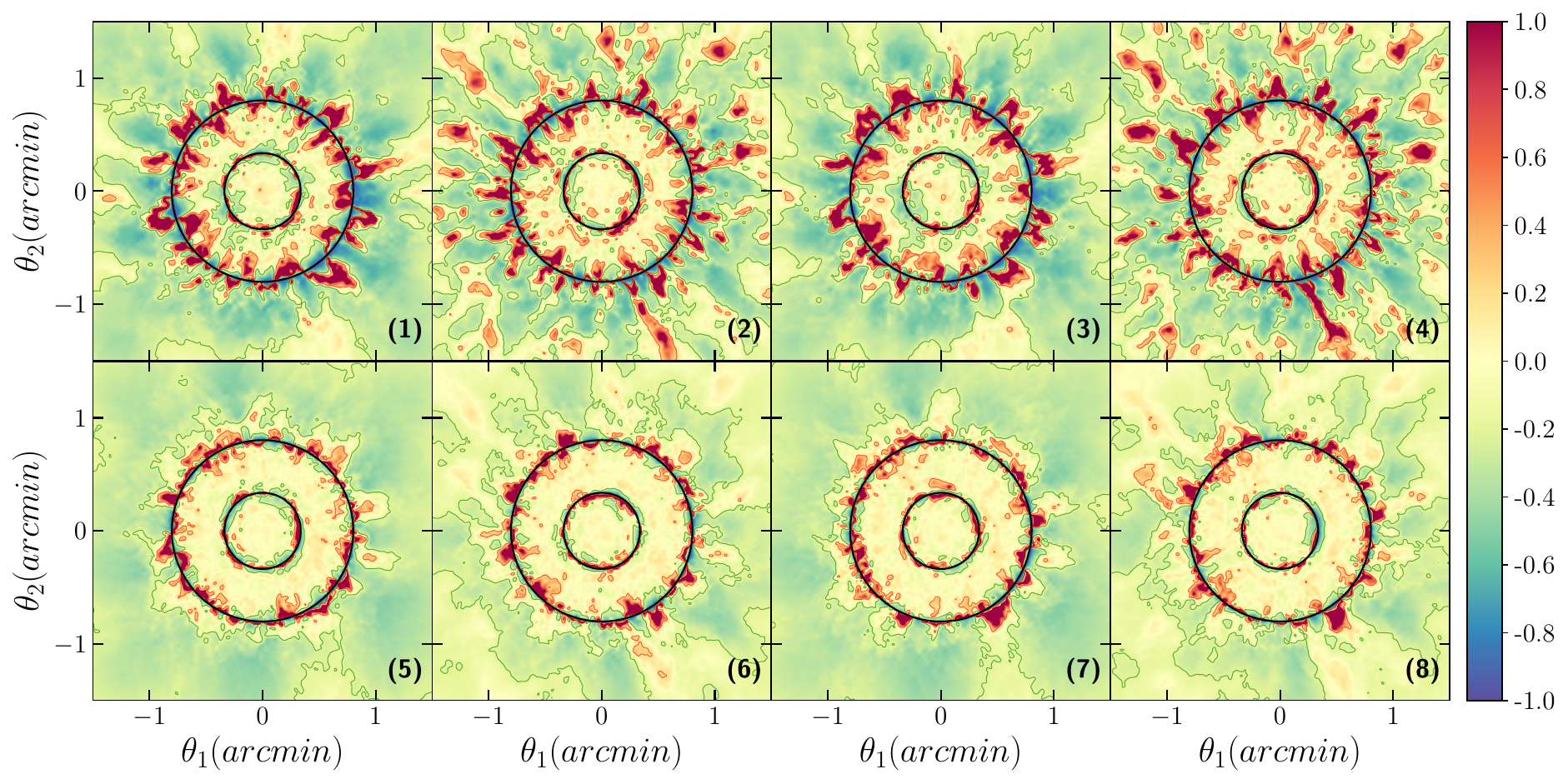}
\caption{Relative difference $\Delta_{r}|\mu|$ between the reconstructed and true magnification maps for the NIS shown in Fig. \ref{fig: nis magnification}. Here, the red and green solid contours correspond to $\Delta_{r}\kappa=0.2$ and $\Delta_{r}\kappa=-0.2$, respectively. The black solid lines depict the true critical curves.}
\label{fig: nis mag comp}
\end{figure*}

\begin{figure*}
\centering
\includegraphics[scale=0.52]{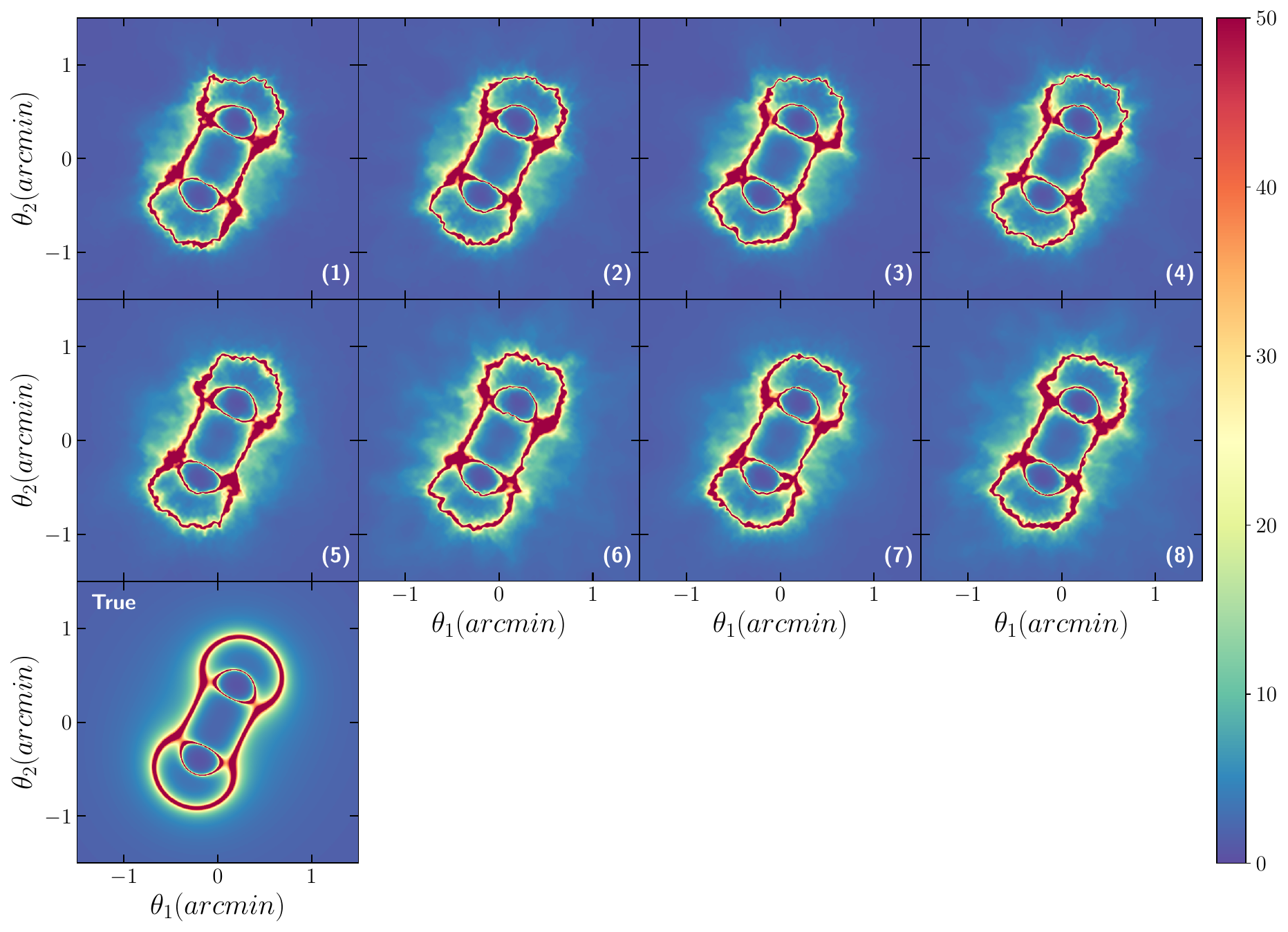}
\caption{Same as in Fig. \ref{fig: nis magnification} but for the 2NIS.}
\label{fig: 2nis magnification}
\end{figure*}
\begin{figure*}
\centering
\includegraphics[scale =0.52]{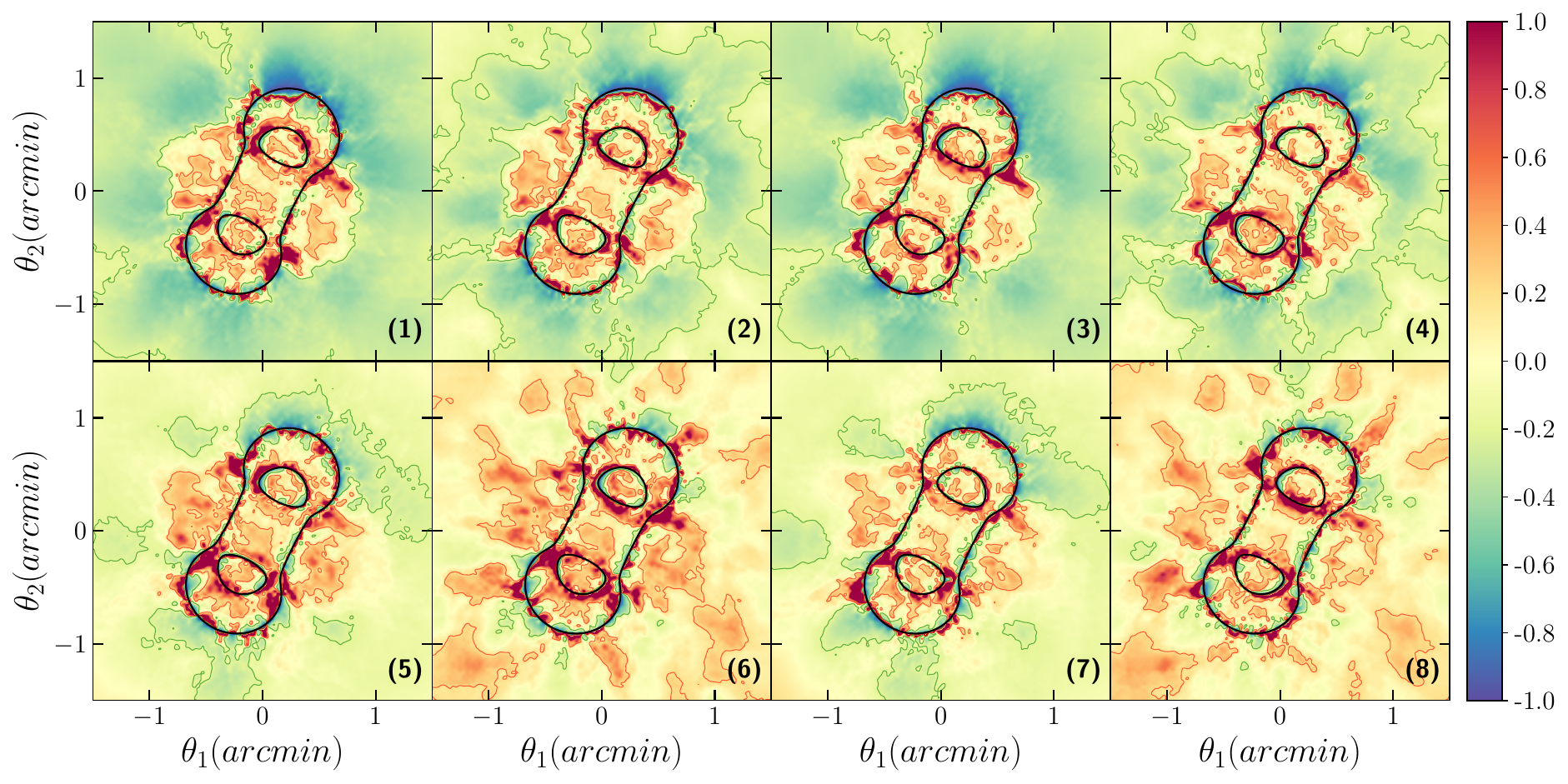}
\caption{Same as in Fig. \ref{fig: nis mag comp} but for the 2NIS.}
\label{fig: 2nis mag comp}
\end{figure*}

With respect to $M$,  which is computed within $0\,arcmin\leq\theta\leq 1.5\,arcmin$, we have a good agreement with the true mass with a relative difference within $|\Delta_{r}M|<0.05$ ($5\%$) for $\theta\leq \theta_{in}\approx 0.74\,arcmin$, and close to $\theta_{out}\approx 0.90\, arcmin$ the curves start to deviate from the true one; except for the orange curves (reconstructions $(6)$ and $(8)$) as it is depicted in Fig. \ref{fig: 2nis mass}. For $\theta>\theta_{out}$ the worst estimation of $M$ is given by reconstructions $(1)$ and $(3)$ (red curves) for which $\Delta_{r}M\approx -0.25$. They are followed by reconstructions $(2)$ and $(4)$ (blue curves), which show an improvement of at most $\sim 7\%$. As expected from the results in $\kappa$, a better scenario is provided by reconstructions $(5)$ and $(7)$ (green curves), as well as $(6)$ and $(8)$ (orange curves), for which the relative difference is at most $|\Delta_{r}M|\approx 0.08$ $(8\%)$ and $|\Delta_{r}M|\approx 0.01$ $(1\%)$, respectively. It is clear that there are not significant differences between Type 1 and Type 2 reconstructions.

Blind reconstructions work incredibly well, particularly if one is interested in the inner region. However, in order to achieve a better understanding of the outskirts, it is worth using an initial guess other than the blind one. The drawback relies on what initial guess we use. For the inner region there are no significant effects if there are enough multiply imaged systems close to the mass peaks, but the outskirts are more sensitive to such changes, where the method presents particular difficulties in scaling down/up to values close to the true $\kappa$, coming from  overestimations/underestimations given by the initial guess.

\subsubsection{Magnification}

\begin{figure}
\centering
\includegraphics[scale =0.5]{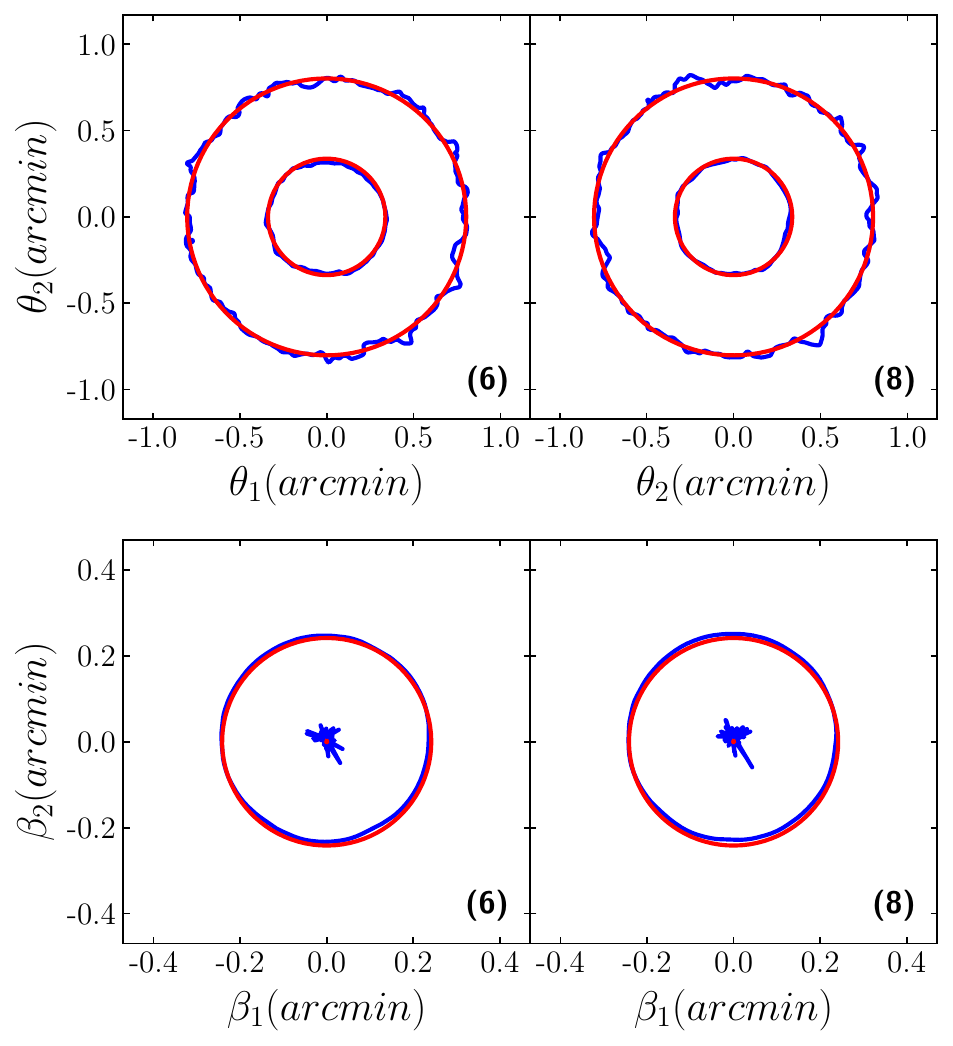}
\caption{Comparison between the reconstructed (blue) and true (red) critical (upper row) and caustic (lower row) curves for the NIS. The curves are given for the reconstructions $(6)$ and $(8)$.}
\label{fig: nis cc}
\end{figure}
\begin{figure}
\centering
\includegraphics[scale =0.5]{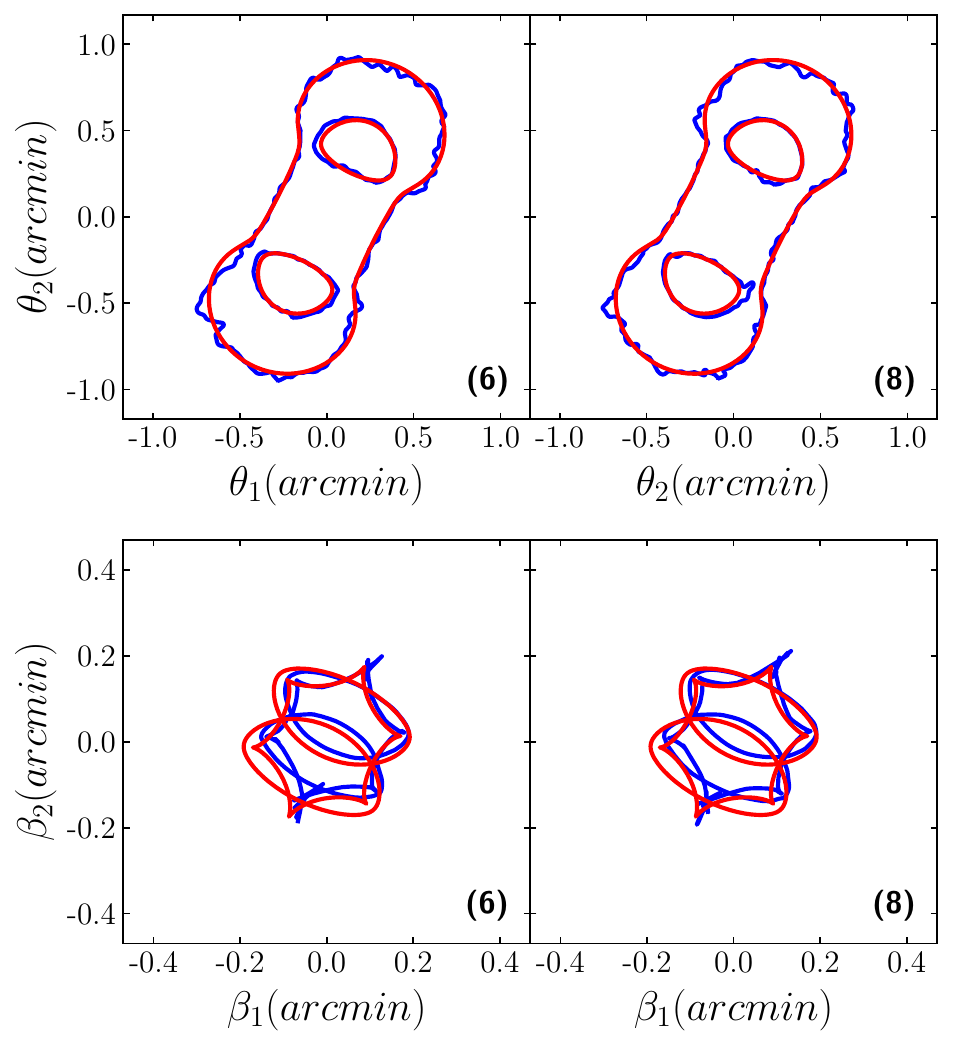}
\caption{Same as in Fig. \ref{fig: nis cc} but for the 2NIS.}
\label{fig: 2nis cc}
\end{figure}

Magnification maps are key elements in the study of galaxy clusters, since they are needed to characterize the galaxy cluster, for example, as a cosmic telescope, since the biggest magnifications occur close to the critical curves, allowing us to study distant objects. Also, they help to identify new multiply imaged systems, as well as to extend those already known. So then, we do not only need the critical curves, but also the magnification itself, close to such curves. The quality of the reconstructed magnification maps has been studied in e.g. \cite{meneghetti2017HFFcomparison}, where they compare the reconstruction of the simulated galaxy clusters, Ares and Hera, given by different methods and approaches, in order to have a better understanding of the results obtained for the HFF clusters. We present our own reconstructions for Ares and Hera in Sec. \ref{sec: reconstruction ares hera}. 

In Fig. \ref{fig: nis magnification} and Fig. \ref{fig: 2nis magnification} we have the magnification maps $|\mu|$ for the NIS and 2NIS, respectively, Those  maps are depicted in concordance with the convergence maps shown in Fig. \ref{fig: nis convergence} and Fig. \ref{fig: 2nis convergence}. 

At critical curves, by definition $\mu$ diverges, so that they are depicted as regions of high $|\mu|$. Due to this behavior, even true deviations from true critical curves are expected to make it difficult to accurately recover $\mu$.

For the NIS, we have that reconstructions $(1)-(4)$ (upper row in Fig. \ref{fig: nis magnification}) produce critical curves that actually follow the form and size of the true ones, but are noisy and quite irregular, particularly the outer or tangential curve. Such irregularities in fact provide a $|\mu|$ map with high error close to the critical curves, as it is depicted in Fig. \ref{fig: nis mag comp} (upper row). Now, by including the smoothing in reconstructions $(5)-(8)$ (middle row in Fig. \ref{fig: nis magnification}) we can see a better recovery of critical curves. They are not as smooth as the true ones, but are less irregular than curves $(1)-(4)$. The improvement is not restricted to the recovery of such critical curves in shape and size, but it extends to the overall estimation of $|\mu|$, particularly when weak lensing is used, as it can be seen for reconstructions $(6)$ and $(8)$ in Fig. \ref{fig: nis mag comp} (lower row). Still, the highest deviations in $|\mu|$ appear close to the critical curves. 

Now, with respect to the 2NIS, in Fig. \ref{fig: 2nis magnification} we can see that there are no significant differences between the critical curves provided by the eight reconstructions. All of them provide accurate reproductions of the critical curves both in shape and size. However, for reconstructions $(5)-(8)$ we get a better reproduction of $|\mu|$ as a whole, as it is depicted in Fig. \ref{fig: 2nis mag comp}. Again, the highest deviations appear close to the critical curves.

Additionally, in Fig. \ref{fig: nis cc} and Fig. \ref{fig: 2nis cc} we can see the critical and caustic curves for the NIS and 2NIS, respectively, given for reconstructions $(6)$ and $(8)$; which have shown the best performance as a hole for both lenses. The critical curves have been computed applying the algorithm discussed in \cite{bartelmann2003numerical}. With respect to the caustic curves, we have a good agreement with the true curves. However, they are affected by the irregularities in the critical curves. This is particularly evident for the NIS, where the tangential critical curve is supposed to be mapped into a point, but instead the reconstructed caustic is wider and more irregular. This effect has to be taken into account when one is interested in predicting the existence of new images or multiply imaged systems, as well as when it comes to verifying the reproduction of the input data. 

It is worth noting that transformation \eqref{eq: potential transformation sourceplane} does not affect the recovery of neither the critical curves nor the convergence map and thus the mass profile, since $\kappa$ and $\gamma$ are invariant under this transformation. Nevertheless, $\boldsymbol{\alpha}$ is not invariant, since $\boldsymbol{\alpha}\to\boldsymbol{\alpha}^{\prime}=\boldsymbol{\alpha}+\boldsymbol{c}$ translates the source plane. It is not an observable, but it results in a translation of the caustic curves. We see that fixing the corresponding three degrees of freedom allows the reconstruction to properly account for the caustic curves, as we have in Fig. \ref{fig: nis cc} and Fig. \ref{fig: 2nis cc} (lower row).

\begin{figure*}
\centering
\includegraphics[scale =0.55]{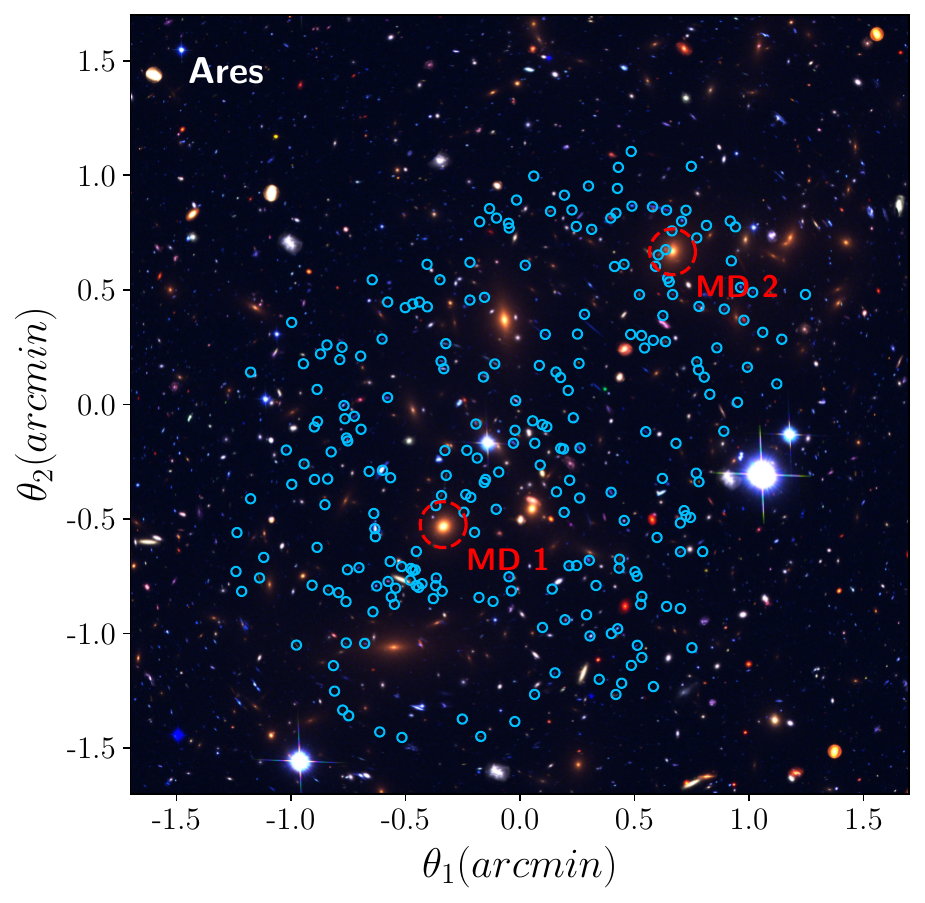}
\includegraphics[scale =0.55]{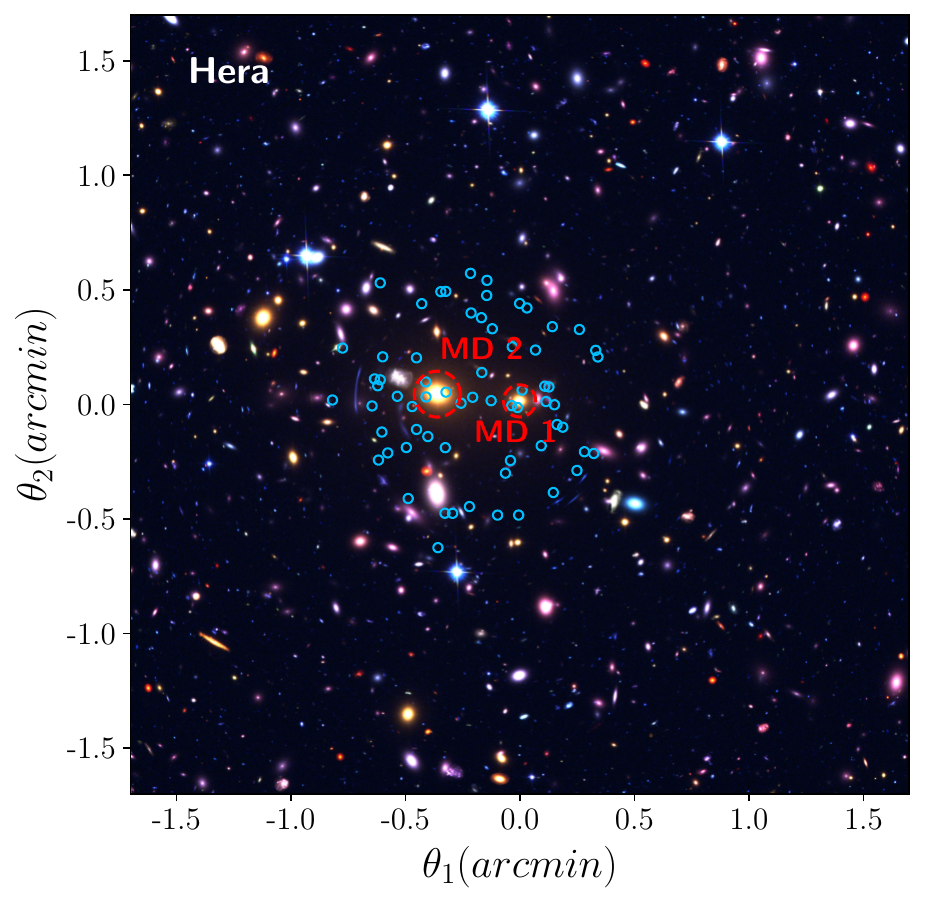}
\caption{Images of Ares (left panel) and Hera (right panel) which have been composed from the F435W, F606W and F814W bands. The blue contours enclose the multiple images, while the red dashed contours enclose the main deflectors. The images cover $3.4\times3.4\, arcmin^2$.}
\label{fig: ares and era}
\end{figure*}

\section{Realistic distributions: Ares and Hera}\label{sec: realistic distributions}

At this point, we have explored under which conditions \relensing\ provides the best performance in its current state, by exploring the reconstructions of the NIS and 2NIS. We move forward in order to explore the validity of the reconstructions provided by \relensing\ in more realistic scenarios. We focus on the simulated clusters Ares and Hera, which are intended to reproduce the complex structures and lensing properties observed in the Hubble Frontier Fields (HFF) clusters.

\begin{figure*}
\centering
\includegraphics[scale =0.75]{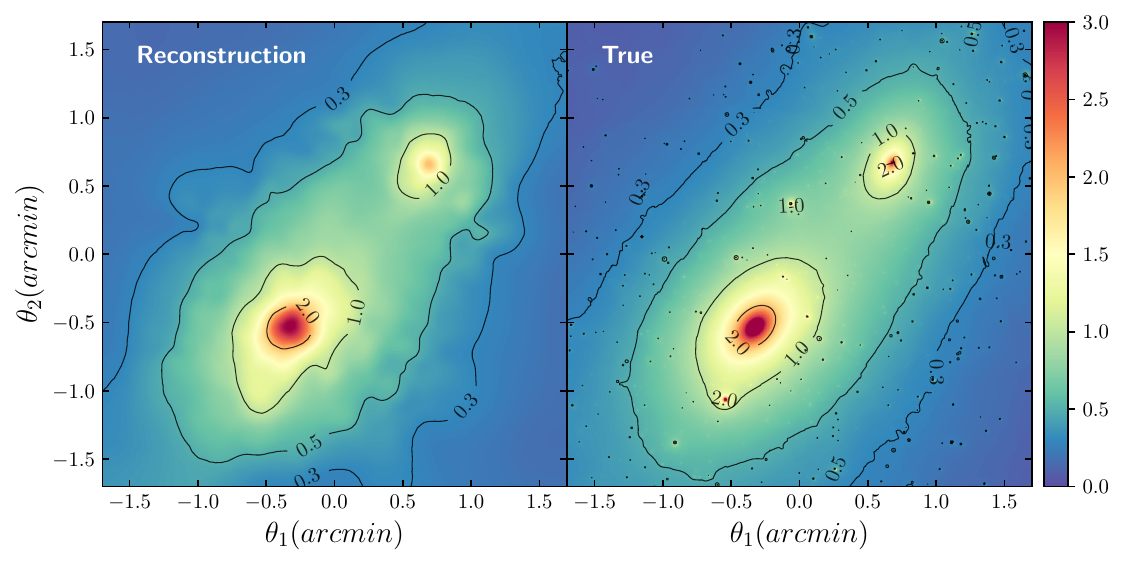}
\caption{Convergence maps ($\kappa$) for Ares, given for a source with redshift $z_r=9$. The reconstruction and true map are shown respectively in the left and right panels.}
\label{fig: ares reco}
\end{figure*}

Ares is shown in Fig. \ref{fig: ares and era} (left panel). It  corresponds to a cluster with redshift $z_l=0.5$, for which a semi-analytical approach was used for its simulation,  under a flat $\Lambda$CDM cosmology with density parameter $\Omega_{m,0} = 0.272$ and Hubble constant $H_0=70.4\,km\,s^{-1}\,Mpc^{-1}$. Ares is characterized by two main distributions or main deflectors (MD) located at $\approx(-0.335,-0.525)\,arcmin $ (MD 1) and $\approx(0.666,0.666)\,arcmin $ (MD 2). Also, it is rich in well defined substructures as it is shown in Fig. \ref{fig: ares reco} (lower panel). 

\begin{figure*}
\centering
\includegraphics[scale =0.75]{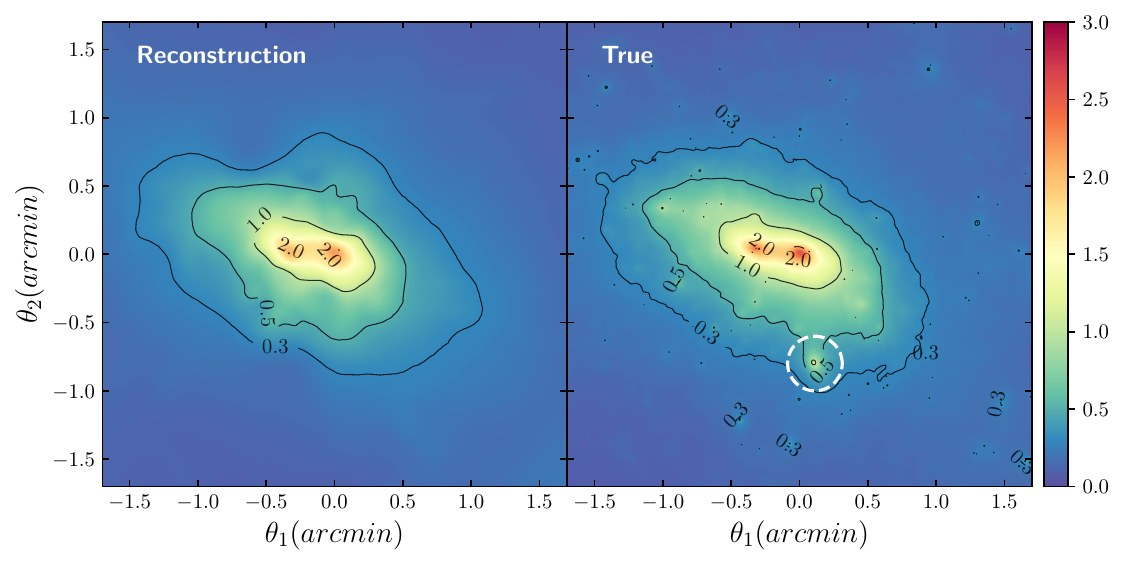}
\caption{Same as in Fig. \ref{fig: ares reco} but for Hera. The white dashed contour encloses the substructure present in the true map of Hera (right panel), and which is not recovered by the reconstruction (left panel).}
\label{fig: hera reco}
\end{figure*}

On the other hand, Hera, with redshift $z_l = 0.507$ is shown in Fig. \ref{fig: ares and era} (right panel). Unlike Ares, the simulation of Hera was carried out using a N-body approach, under a flat $\Lambda$CDM cosmology with density parameters $\Omega_{m,0} = 0.24$, $\Omega_{b,0} = 0.04$ and Hubble constant $H_0=72\,km\,s^{-1}\,Mpc^{-1}$. Like Ares, Hera shows two prominent peaks corresponding to the MD, located at $\approx(-0.001,0.015)\,arcmin $ (MD 1) and $\approx(-0.361,0.045)\,arcmin $ (MD 2). It also exhibits a clear substructure at $\approx(0.113,-0.883)\,arcmin$, whose influence appears in the convergence map Fig. \ref{fig: hera reco} (lower panel), as well as in the magnification map Fig. \ref{fig: hera magnification} (lower panel).

See \cite{meneghetti2017HFFcomparison} and the references therein for further details on how the simulations of Ares and Hera were carried out, as well as their lensing properties.   

\subsection{Reconstruction}\label{sec: reconstruction ares hera}
For Ares, in the strong regime, there is available a catalogue that consists of 85/242 systems/images. This amount of constraints exceeds those found e.g in the HFF clusters, for which the number of constrains used/identified for their reconstructions have been e.g. 60/188 systems/images in \cite{mahler2018a2744} for Abell 2744,  45/138 systems/images in \cite{lagattuta2019a370} for Abell 370, 19/52 systems/images in \cite{karman2017as1063} for Abell S1063,  48/138 systems/images in \cite{bergamini2021macsj0416, vanzella2021macsj0416} for MACS J0416.1- 2403 (MACS 0416), 60/165 systems/images in \cite{limousin2016macs0717} for MACS J0717.5+ 3745 (MACS 0717), and 45/143 systems/images in \cite{jauzac2016macs1149} for MACS J1149.5+ 2223 (MACS 1149). The number of systems/images actually used for the reconstructions is usually smaller, depending on the uncertainty in their identification. For the weak regime the ellipticities redshifts are not available. Hence, we only use the strong regime for Ares reconstructions. 

In the case of Hera, for the strong regime we have fewer constraints, being  19/65 systems/images. The impact on the number of constraints is noticeable in the reconstruction, as we discuss below. Here, for the weak regime the redshifts are available, therefore, for Hera we use both strong and weak regimes to preform the reconstruction. The reconstruction takes place in a region of $3.4\times 3.4\,arcmin^2$ for which we have 123 sources with their ellipticity and redshift.

\relensing\ approaches the refinement of the grid with respect to the main deflectors. So then, in order to identify what cluster members can be considered as main deflectors, one can simply apply a quick blind reconstruction, which will show the peaks of mass corresponding to those main deflectors. Following this process, we identified the two main deflectors present in Ares, as well as in Hera. Then, we estimated their location, which were aforementioned.

Now, with respect to the reconstructions, for each cluster we take as initial guess a NIE as well as a 2NIE; since both cluster have two main deflectors. Additionally, from our previous experience  with the NIS and 2NIS we have that Type 1 and Type 2 reconstructions produce similar results. Hence, it is best to let them work together.

Considering the input potential and the type of reconstruction,  we are left with four combinations to carry out the reconstructions, namely NIE + Type 1, NIE + Type 2, 2NIE + Type 1, and 2NIE + Type 2. For each combination we perform $10$ realizations of the reconstructions, which give us $40$ reconstructions in total. The results we show in this section correspond to the average of  those $40$ reconstructions. We do this for Ares and Hera.

The conditions to perform the reconstructions are the same for Areas and Hera. For all realizations we consider the smoothing. We consider an initial grid of size $n_x\times n_y$ nodes over a region of $3.4\times3.4 \,arcmin^2$; which covers the region shown in Fig. \ref{fig: ares and era}. Since we are working on a square region, we consider $n_x=n_y$, which are randomly selected between $20-25$ nodes. Likewise, $\eta_{\kappa}=\eta_{\gamma}$ are randomly selected between $100-200$.

With respect to the refinement process, we add $N_d$ new nodes for each refinement for each main deflector, which is randomly between $300-500$. Such nodes are distributed with $\sigma_d = 1\,arcmin$, and $R_d = 2\,arcmin$. The adaptive refinement was done with $r_d = 0.2\,arcmin$ and $u=0.8$. For the finite differences we use $Q=25$ nodes. 

We have used almost the same parameters that we used for the NIS and 2NIS, except for $n_x$, $n_y$, $\eta_{\kappa}$, $\eta_{\gamma}$, and $N_d$, which were drawn from a uniform distribution defined in the given ranges. The selection took place for each realization of the reconstruction.

\begin{figure*}
\centering
\includegraphics[scale =0.85]{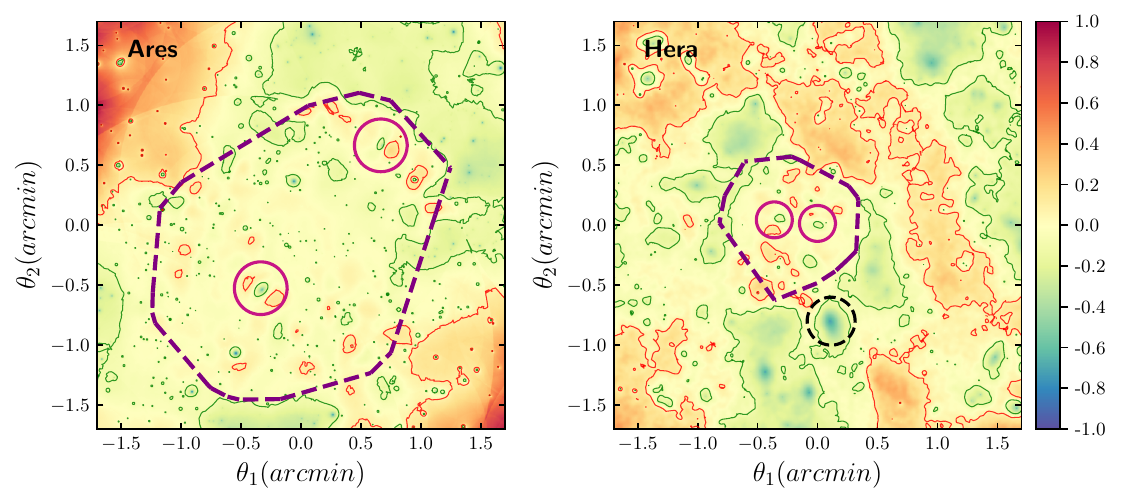}
\caption{Relative difference $\Delta_{r}\kappa$ between the reconstructed and true convergence maps for Ares (left panel), as well as for Hera (right panel). Here, the red and green solid contours correspond to $\Delta_{r}\kappa=0.1$ and $\Delta_{r}\kappa=-0.1$, respectively. The purple dashed contour delimits the region within which we have multiple images available. Additionally, the solid violet contours enclose the main deflectors. Here, the black dashed contour encloses the substructure present in Hera (left panel).}
\label{fig: reco_comp}
\end{figure*}

\begin{figure}
\centering
\includegraphics[scale =0.52]{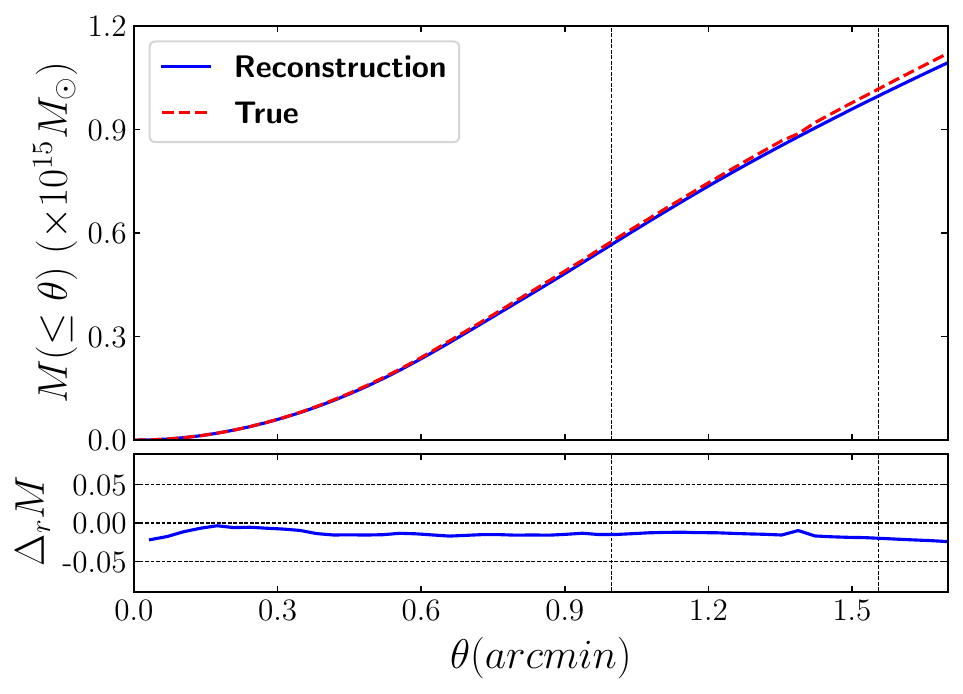}
\caption{Mass for Ares within an angular radius $\theta$, centered at $\boldsymbol{\theta} = (0,0)\,arcmin$ (upper panel), and the relative difference $\Delta_{r}M$ comparing the reconstructed and true mass (lower panel). The vertical dashed lines represent the inner and outer radius of the inner region (purple dashed contour in Fig. \ref{fig: reco_comp} left). The horizontal dashed lines  enclose $|\Delta_{r}M|\leq 0.05$.}
\label{fig: ares mass}
\end{figure}

\begin{figure}
\centering
\includegraphics[scale =0.52]{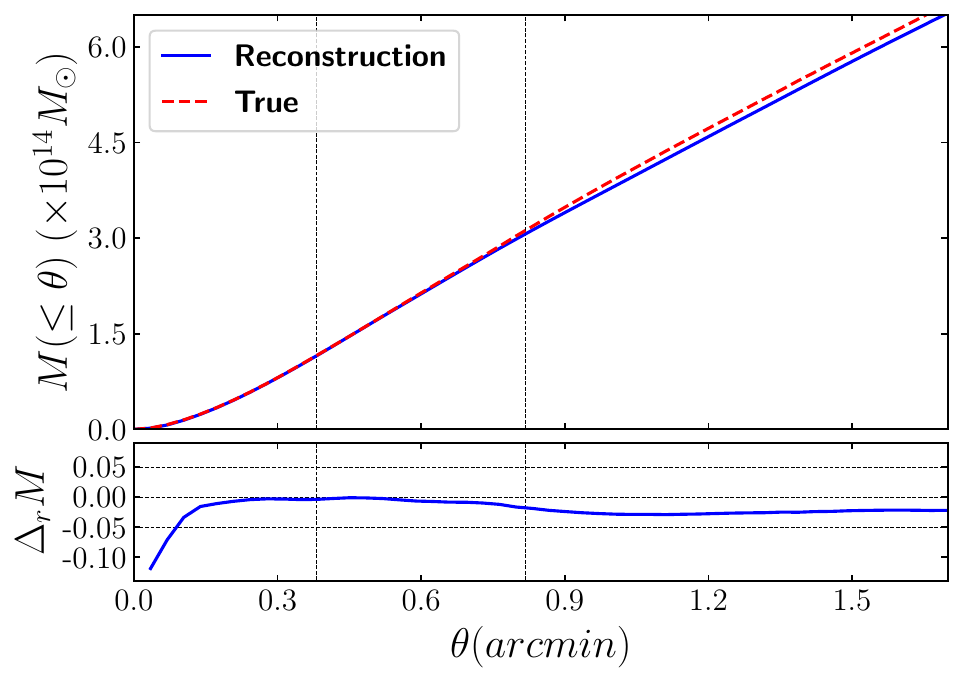}
\caption{Same as in Fig. \ref{fig: ares mass} but for Hera.}
\label{fig: hera mass}
\end{figure}

\subsubsection{Convergence}
In Fig. \ref{fig: ares reco} we find the reconstructed (left panel) and true (right panel)$\kappa$ maps for Ares. It is clear that our reconstruction retrieves successfully the two MD, as well as the general morphological characteristics found in Ares. However, the reconstructions cannot account for the rich substructure present in Ares. 

Within the inner region we have that most of such region satisfies $|\Delta_{r}\kappa|<0.1\,(10\%)$ as it is depicted in Fig. \ref{fig: reco_comp} (left panel), where we also can find few regions for which $0.1\,(10\%)\leq\Delta_{r}\kappa\leq0.2\,(20\%)$ (red contour) and $-0.2\,(20\%)\leq\Delta_{r}\kappa\leq-0.1\,(10\%)$ (green contour). Beyond that, we have $|\Delta_{r}\kappa|>0.2\,(20\%)$ at those locations with substructure. Outside the inner region the quality of the reconstruction decreases as expected, since we are using strong lensing only. The discrepancy is higher towards the upper left and lower right corners, due to the lack of observations there, as we can see in Fig. \ref{fig: ares and era} (left panel). 

For Hera, we can see in Fig. \ref{fig: hera reco} that our reconstruction (left panel) recovers both MD, as well as the general shape and size shown by Hera. However, they do not exhibit the substructure (enclosed by the white dashed curve) found in the true profile (right panel). In Fig.\ref{fig: reco_comp} (right panel) we can see that such substructure (enclosed by the black dashed curve) lies outside the inner region, so that there are not multiple images there that allow \relensing\ to account for it. Therefore, the absence of this substructure is due to the lack of data there, and not a consequence of the reconstruction method used. This behaviour is consistent with the reconstructions discussed in \cite{meneghetti2017HFFcomparison}. Now, similarly to Ares, we can see that $|\Delta_{r}\kappa|<0.1\,(10\%)$, for most of the inner region,  with few regions where $0.1\,(10\%)\leq\Delta_{r}\kappa\leq0.2\,(20\%)$ (red contour) and $-0.2\,(20\%)\leq\Delta_{r}\kappa\leq-0.1\,(10\%)$ (green contour). For the substructure we have $\Delta_{r}\kappa>-0.3\,(30\%)$.

In addition, for both Ares and Hera we can see in Fig. \ref{fig: reco_comp} that our reconstructions present a deviation $|\Delta_{r}\kappa|\gtrapprox 0.1\,(10\%)$ around the MD (violet solid contours), where the underestimations are present at the exact location of the MD (green contours). 

With respect to $M$, it is computed in the range $0\,arcmin\leq\theta\leq 1.7\,arcmin$.  For Ares, the maximum relative difference that our reconstruction produces is $|\Delta_{r}M|\approx 0.024\,(2.4\%)$, which appears close to the center of the distribution. The deviation from the true $M$ is stable even between $\theta_{in}\approx 1\,arcmin \leq\theta\leq\theta_{out}\approx 1.56\, arcmin$, and it starts to deviate beyond $\theta_{out}$, as we saw for the NIS and 2NIS. 

For Hera, we can see in Fig. \ref{fig: hera mass} that the deviation from the true $M$ gets its higher value $|\Delta_{r}\kappa|\approx 0.12\,(12\%)$ for $\theta<0.2\,arcmin$, which approximately coincides with the mass estimation for MD 1; since it is located close to the image center. This behaviour is expected from how the convergence behaves close to the MD, where, as we mentioned above, it presents a underestimation $|\Delta_{r}\kappa|\gtrapprox 0.1\,(10\%)$, which translates in a underestimation in $M$ with the same range of error. Beyond such underestimation, the difference decreases rapidly, achieving $|\Delta_{r}\kappa|\approx 0\,(0\%)$. Between $\theta_{in}\approx 0.38\,arcmin \leq\theta\leq\theta_{out}\approx 0.82\, arcmin$  again $M$ starts to deviate from its true value, getting up to $|\Delta_{r}M|\approx 0.05\, (5\%)$ beyond $\theta_{out}$. 

Among the reconstructions discussed in \cite{meneghetti2017HFFcomparison}, we compare our results to those reconstructions produced with \SWUnited, since \SWUnited\ and \relensing\ are based on the same principles. In \cite{meneghetti2017HFFcomparison} they labeled the reconstructions produced with \SWUnited\ as Bradac-Hoag models. Here, we will refer to them as BHm.

The $\kappa$ maps given by BHm (for a source with redshift $z_r=9$) for Ares and Hera are depicted respectively in Fig. 7 (upper let panel) and Fig. 8 (upper left panel) in \cite{meneghetti2017HFFcomparison}. One can see that for Ares their reconstruction shows a more irregular shape, and noisier contour curves than our reconstruction. In the case of Hera, BHm is smoother than the one given for Ares, however, their reconstruction exhibits a substructure close to MD 1 that is not present in the true distribution. This extra substructure is not present in our reconstructions, as it is clear from Fig. \ref{fig: hera reco}. For BHm as well, the substructure present in Hera was not recovered.

\begin{figure*}
\centering
\includegraphics[scale =0.75]{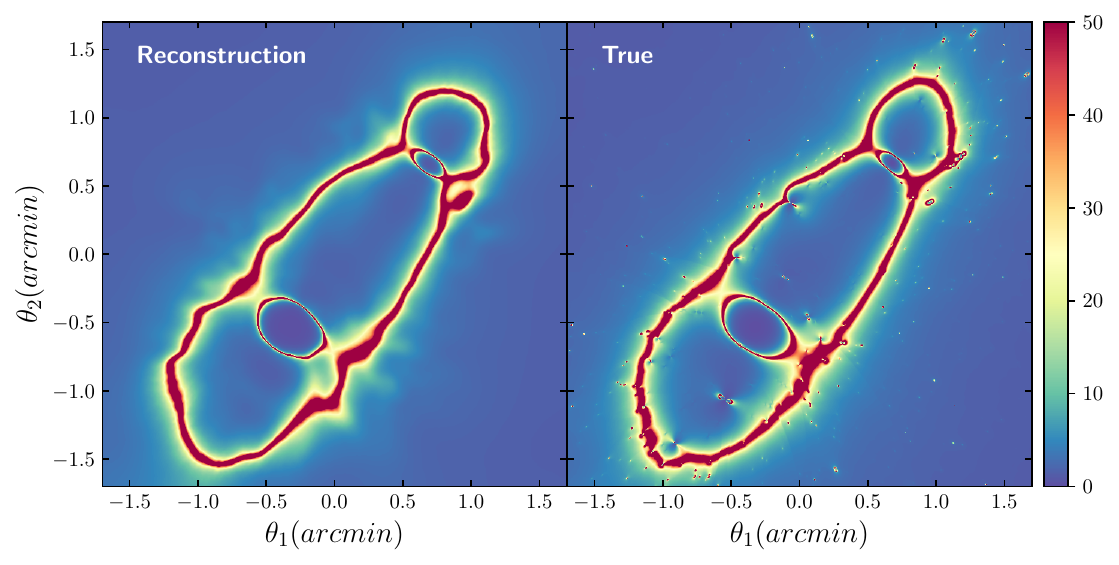}
\caption{Magnification maps ($|\mu|$) for Ares, given for a source with redshift $z_r=9$. They correspond to the maps shown in Fig. \ref{fig: ares reco}. The true map is shown in the left panel.}
\label{fig: ares magnification}
\end{figure*}

\begin{figure*}
\centering
\includegraphics[scale =0.75]{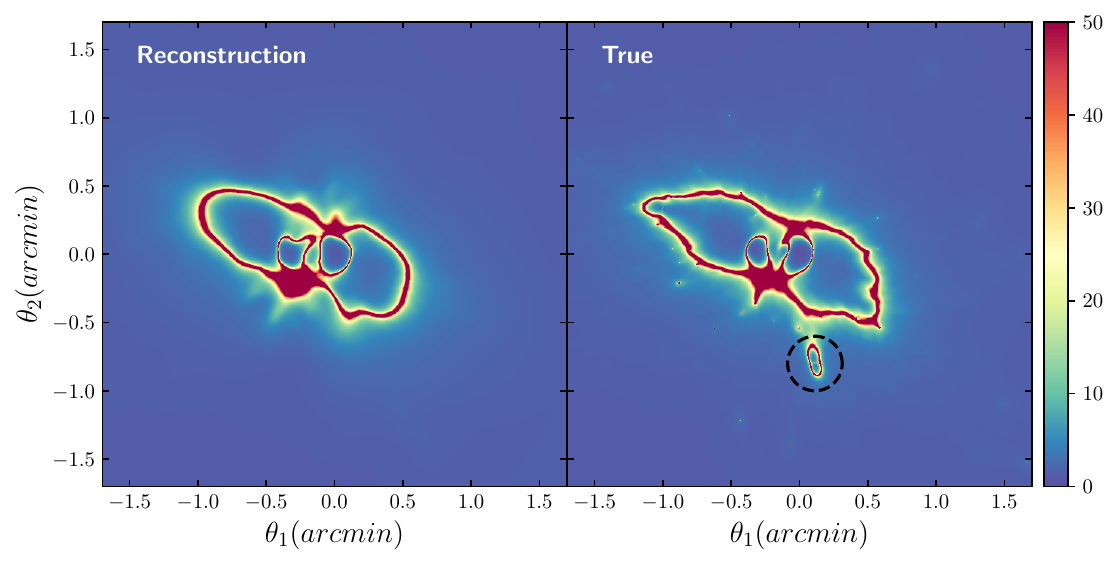}
\caption{Same as in Fig. \ref{fig: ares magnification} but for Hera. Here, the black dashed contour encloses the critical curve produced by the substructure present in Hera.}
\label{fig: hera magnification}
\end{figure*}

\begin{figure*}
\centering
\includegraphics[scale =0.85]{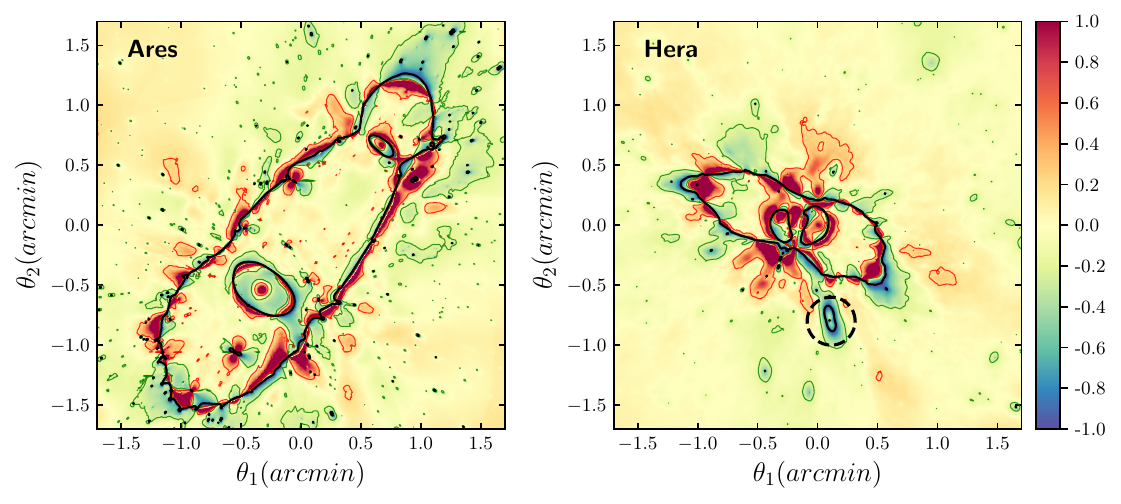}
\caption{Relative difference $\Delta_{r}|\mu|$ between the reconstructed and true magnification maps for Ares (left panel) and Hera (right panel). The red and green solid contours correspond to $\Delta_{r}\kappa=0.2$ and $\Delta_{r}\kappa=-0.2$, respectively. The black solid lines depict the true critical curves. Additionally, the black dashed contour encloses the critical curve produced by the substructure present in Hera.}
\label{fig: reco_mag_comp}
\end{figure*}

\begin{figure*}
\centering
\includegraphics[scale =0.8]{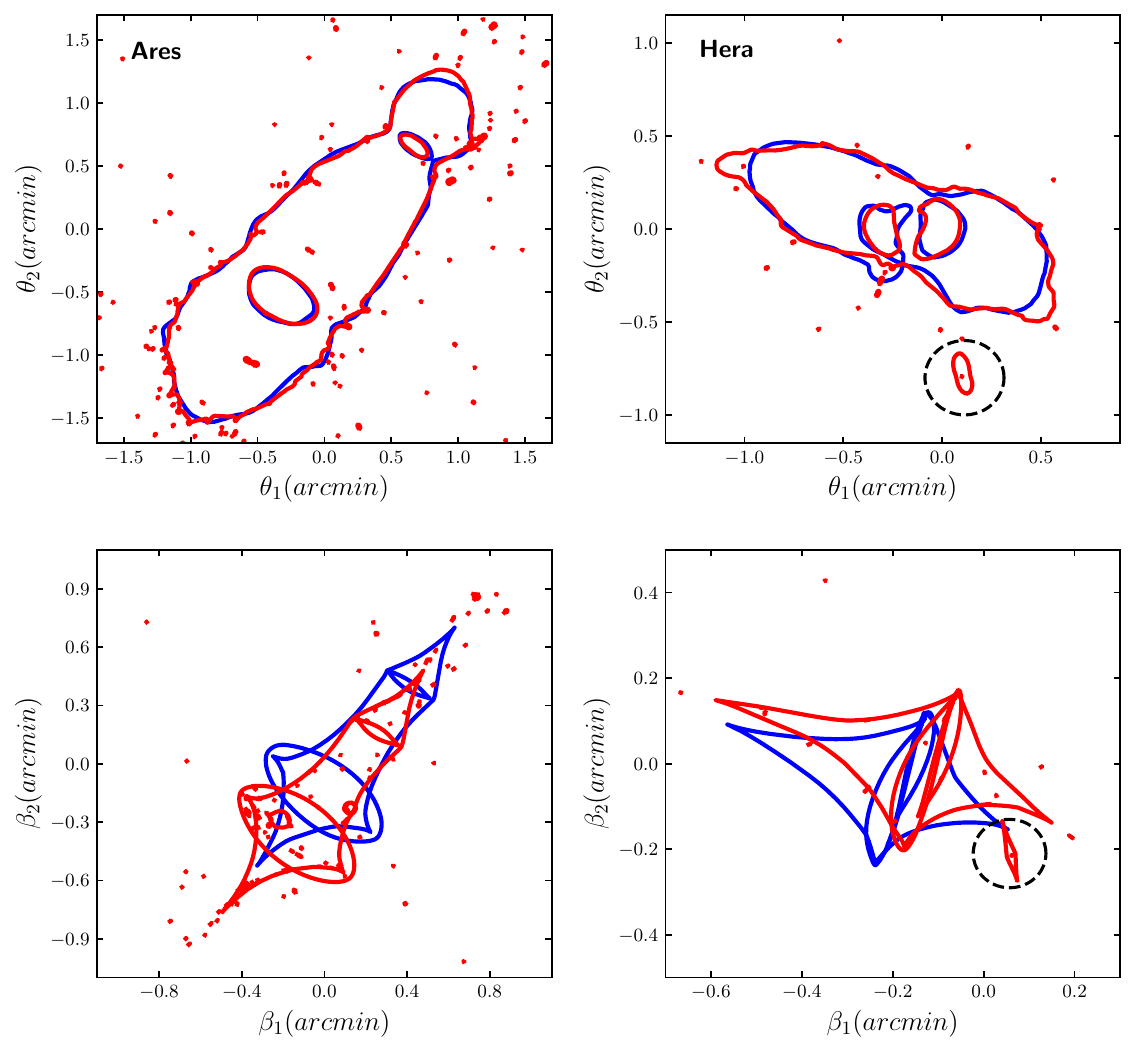}
\caption{Comparison between the reconstructed (blue) and true (red) critical (upper row) and caustic (lower row) curves for Ares (left panels) and Hera (right panels). The black dashed contour encloses the critical (upper right panel) and caustic (lower right panel) curves produced by the substructure present in Hera.}
\label{fig: ares_hera cc}
\end{figure*}

\subsubsection{Magnification}
In Fig. \ref{fig: ares magnification} we can see the magnification maps for Ares, where the reconstruction is shown in the left panel, and the true map in the right panel. The external curve of high $|\mu|$ exhibits an elongated shape, within which there is one internal curve surrounding each MD. The main difference appears close to the MD 2, where we get an overestimation of $|\mu|$. Such overestimation coincides with a true critical curve, as one can notice by direct comparison between the true (red) and reconstructed (blue) critical curves in Fig. \ref{fig: ares_hera cc} (upper left panel). Despite this overestimation, we are not able to retrieve such critical curve. It is worth mentioning that it is possible to reproduce such critical curve by forcing the reconstruction parameters, where its presence is more frequent for Type 2 reconstructions. Nonetheless, in practice we are not aware of these details, so that the best we can do is to use the parameters that provide the best overall outcome. 

It is clear that our reconstruction recovers the orientation, shape and size of such critical curves with great accuracy. Additionally, the critical curves corresponding to the substructure cannot be recovered, since it is difficult to account for such detailed peaks from the limited information provided by strong lensing. 

For Hera, its magnification maps are shown in Fig. \ref{fig: hera magnification}. In general terms, our reconstruction is capable of recovering the main features shown by the true map (Fig. \ref{fig: hera magnification} left panel), except for the curve (surrounded by the white dashed contour) produced by the substructure present in Hera, which our reconstruction fail to account for. As we discussed above, it is not possible to account for such substructure, due to the lack of observations there. Besides such curve, the external critical curve exhibits an elongated shape with one inner curve surrounding each MD; similarly to Ares. Fig. \ref{fig: ares_hera cc} (upper right panel) shows the direct comparison between true (red) and reconstructed (blue) critical curves. It is clear that the shape, and size of the main curves are reproduced by our reconstruction.

Since for Hera we have fewer systems/images than for Ares, it is expected for us to obtain less accurate curves. This can be seen in the external curve, where at its ends the reconstructions fail to accurately reproduce its shape. This becomes more evident in the critical curves, as it is depicted in Fig. \ref{fig: ares_hera cc} (upper left panel). Nonetheless, our results are incredibly accurate, considering that the only assumption that we have made with respect to the cluster itself relies on the NIE and 2NIE used as input. Beyond that, the reconstructions just adapt to the observations.      

From Fig. \ref{fig: reco_mag_comp} it is clear again that the higher difference between the true and reconstructed $|\mu|$ maps is produced close to the true critical curves (solid black lines), where the deviation from the true curves shows the overestimation in $|\mu|$ associated with the presence of the reconstructed critical curves. This is more evident for Hera.

Now, moving to the source plane, in Fig. \ref{fig: ares_hera cc} we have the true (red) and predicted (blue) caustic curves for Ares (lower left panel) and Hera (lower right panel). We successfully recover the overall morphological characteristics present in the caustic curves, except for those curves associated with substructure. Our curves are deviated from the true ones, which is a consequence of \eqref{eq: potential transformation sourceplane}. 

Now, for the $|\mu|$ maps given by BHm, which are shown for Ares and Hera respectively in Fig. 19 (upper left panel) and Fig. 20 (upper left panel) in \cite{meneghetti2017HFFcomparison}, it is clear that the irregularities present in Ares produce equally irregular critical curves. BHm is capable of reproducing the curves around the MD, and the external curve follows the orientation of the true curve. Nonetheless, the external curve produces several regions of high magnification that do not follow the true map. In contrast, our reconstruction has produced smoother curves which present less deviations from true critical curves. On the other hand, the BHm of Hera shows a critical curve related to the substructure present in their reconstruction, which is absent in the true map. Since our reconstruction does not present such substructure, it lacks that critical curve. 

Lastly, we consider the root-mean-square (rms) on the lens plane, define as
\begin{equation}\label{eq: rms}
\Delta_{rms} = \sqrt{\dfrac{1}{\tilde{N}_{\text{img}}}\sum_{i=1}^{N_s}\sum^{\tilde{N}_{i}}_{n=1}\Big\vert\boldsymbol{\theta}_{in}-\boldsymbol{\theta}^{ \prime}_{in}\Big\vert^2},
\end{equation}
where $\tilde{N}_i$ is the number of reproduced images corresponding to the $i$-th multiply imaged system, and $\tilde{N}_{\text{img}}=\sum_{i}^{N_s}\tilde{N}_i$ is the total number of multiple images recovered from the reconstruction. Recall that $N_s$ is the number of multiple imaged systems used in the reconstruction, as well, $\boldsymbol{\theta}_{in}$ and $\boldsymbol{\theta}^{\prime}_{in}$ are the $n$-th reconstructed and true angular position of the $i$-th multiply imaged system, respectively. In general, $\tilde{N}_{\text{img}}\neq N_{\text{img}}$, since it is possible to get an excess or lack of multiple images. In case of getting an excess of multiple images, we discard those which are outside a given neighborhood with respect to the closest true image.

A small rms is an indication of how well the reconstruction reproduces the input image positions. However, in practice it is perhaps more important to get a small prediction rms (prms), which is computed for a set of multiply imaged systems which were not considered as input data.  Now, take into account that, as it is discussed in e.g \cite{2018liliya_grale}, a small rms by itself is a necessary but not sufficient criteria to qualify the overall quality of the reconstruction, particularly since it may be the indication of an overfitting of the intrinsic noise in the data. Thus, it becomes necessary to get small values for both, rms and prms.  For further details in this regard see e.g \cite{2013statisticallearnig}.  We left the study of the prediction power of \relensing\ for another paper.

For Ares we have $\Delta_{rms} \approx0.17\,arcsec$ with $\tilde{N}_{\text{img}}=240$ out of the total $N_{\text{img}}=242$. Meanwhile, for Hera we get  $\Delta_{rms} \approx0.16\,arcsec$ with $\tilde{N}_{\text{img}}=N_{\text{img}}=65$.  Considering the rms aforementioned, as well as the reproduction of the different properties of Ares and Hera discussed up to here, we are confident to say that \relensing\ is capable of producing accurate reproduction of the mass distributions of galaxy clusters, and their properties as gravitational lens.

\section{Summary and conclusions}\label{sec: conclusions}
In this work we describe and test a free-form method which makes use of gravitational lensing in order to produce an estimation of the mass profile of galaxy clusters, as well as its properties as a gravitational lens. In addition, we note that this approach does not consider that mass traces light; commonly assumed in parametric methods. This approach is an extension of the method presented in \cite{bradac2005reco, bradac2009cosmictelescope}, which at the same time is an extension of the work discussed in \cite{bartelmann1996freeform}. 

Here, we use an irregular and adaptive grid that is intended to produce a higher resolution around those mass peaks (main deflectors) responsible for most of the strong lensing  effect, as well as around the multiple images. Additionally, we have opted for an alternative finite difference approach (generalized finite difference). Moreover, we include two different ways of computing the penalty function $\chi_{s}^2$, which we have named Type 1 and Type 2.

We start by testing our approach on two simple distributions; a NIS and 2NIS, in order to explore the set of input parameters that take us to reliable results. With them, we also explore the impact that the input deflection potential has on the reconstructions, where we use a blind reconstruction and a NIE. Within the region delimited by the available multiple images (inner region) there are no major differences. However, for blind reconstructions it is difficult to scale up outside the inner region. The introduction of weak lensing improves the reconstruction on such region, but there is still a lack of mass towards the outskirts. The NIE helps to extend the reconstruction outside the inner region along with weak lensing, where for the 2NIS we get an improvement in the mass estimation of $\sim 7\%$. For $\kappa$ most of the region presents an estimation that lies below $\sim 20\%$, with just a few sectors mainly close to the boundaries where we get higher values. Thus, the choice of $\psi^{(0)}$ does not affect significantly the reconstruction where strong lensing is dominant. However, outside such region the reconstruction is more susceptible to $\psi^{(0)}$. If $\psi^{(0)}$ corresponds to an overestimation of $\kappa$ that moves away from the true map, it becomes difficult for the reconstruction to scale down towards the true value. The opposite effect occurs in blind reconstructions. 

We also show that by recomputing the deflection potential from the weight of every node; i.e by using \eqref{eq: finite diff pot}, we are able to smooth the shape of convergence and magnification maps, which otherwise tend to be quite irregular and noisy. For instance, the overall relative difference $\Delta_{r}\kappa$ and $\Delta_{r}\mu$ improves. We refer to this process as smoothing. 

This smoothing improves the power of this free-form approach, which we also test on Ares and Hera \cite{meneghetti2017HFFcomparison}, two simulated galaxy clusters, which provide a more realistic framework to prove the reliability of our reconstructions. Our results show an improvement with respect to e.g. BHm (Bradac-Hoag models) in \cite{meneghetti2017HFFcomparison}, where we get less irregular profiles and magnification maps, which turns into a more accurate reconstruction of these distributions. 

Among the reconstructions discussed in this work, we have that within the inner region the reconstructions mainly satisfy $|\Delta_r{\kappa}|<0.1\,(10\%)$, with $0.1\,(10\%)<\Delta_r{\kappa}<0.2\,(20\%)$ mostly close to inner region boundary. Higher values are obtained most likely where small substructures are located, like in Ares. When weak lensing is included, the reconstruction provides a more accurate estimation of the mass profile outside the inner region. For $|\mu|$, the higher discrepancies appear close to the critical curves, as expected due to their nature. The morphological properties of the critical and caustic curves do not show significant changes when weak lensing is also used in the reconstruction.  

Our reconstructions have shown to be capable of recovering the orientation, shape, and size of critical curves, despite the irregularities present in the reconstructions. We have found a way of reducing such irregularities by means of the smoothing. Nevertheless, since critical curves are sensitive to variations by nature, they might produce high deviations in $|\mu|$; even if such variations are small. This is a problem that requires further improvement.      

We have seen that Type 1 and Type 2 reconstructions produce similar results, except for some minor differences that are related to the input parameters. Therefore, it is best to use both types of reconstruction for a given galaxy cluster, which is done by taking several realizations of the reconstruction for each type of reconstruction combined with different initial deflection potentials (if convenient). If this is done, it is also convenient to use random seeds for some input parameters. Then, by averaging such realizations, the result is even more accurate. This process becomes convenient considering the random nature of the grid along the reconstruction.  

We followed this approach for Ares and Hera, where we used the NIE and 2NIE as an initial guess, so that we averaged $40$ realizations of the reconstructions for each cluster. As a result, we got a rms on the lens plane of about $0.17\,arcsec$ and $0.16\,arcsec$ for Ares and Hera. Also, we were able to recover $240$ of the $242$ images used in Ares, and all the images used in Hera. 

\relensing\ has shown an incredible capability in terms of producing an accurate estimation of the mass profile of galaxy clusters, along with their properties as gravitational lens. Therefore, with this work we expect to provide an accessible package written in \python, focused on the characterization of galaxy clusters by using gravitational lensing. This provides users with a wider range of options to choose from, which becomes indispensable in order to validate physical results. 

In this regard, the application of \relensing\ to real galaxy clusters, such as those from e.g. HFF, is naturally the next step (this is a work in progress). Also, with the expected increase in quantity and quality of observations from upcoming facilities, we plan to explore how \relensing\ behaves depending on the number of multiply imaged systems, and multiple images available. Additionally, as already discussed in \cite{cain2016flexion}, flexion allows us to get information from substructure where strong lensing is not dominant. So that it becomes interesting to explore the impact the smoothing has on reconstructions when flexion is included. Moreover, it is indispensable to test the prediction power of \relensing\, particularly in the context of predicting the existence of multiple images, which is of great importance for strong lensing studies.

\section*{Acknowledgements}
L. Casta\~neda was supported by Patrimonio Autónomo - Fondo Nacional de Financiamiento para la Ciencia, la Tecnología y la Innovación Francisco José de Caldas (MINCIENCIAS - COLOMBIA) Grant No. 110685269447 RC-80740-465-2020, projects 69723.

We thank A. Plazas for his comments on our work, and the suggestion of applying our approach to Ares and Hera. We also appreciate the insights of L. A. Garcia, J. Liesenborgs, and L.L.R. Williams, which have been useful to improve our manuscript. Additionally, we thank M. Meneghetti and P. Natarajan for making publicly available the simulations of Ares and Hera through \href{http://pico.oabo.inaf.it/~massimo/Public/FF/index.html}{FF-SIMS}.

\relensing\ has been possible thanks to \numpy\ (\cite{harris2020array}), \scipy\ (\cite{2020SciPy-NMeth}), \numba\ (\cite{lam2015numba}), \matplotlib\ (\cite{Hunter_2007}), and \astropy\ (\cite{astropy:2013, astropy:2018}).

\section*{Data Availability}
The data discussed in this work will be shared upon request to the authors. We plan to make \relensing\ available through \github\ in the near future.



\bibliographystyle{mnras}
\bibliography{references} 




\appendix




\bsp	
\label{lastpage}
\end{document}